%% file: WSS_GGSP_arx_v1.tex
\pdfoutput=1
\documentclass[10pt, twocolumn, final]{IEEEtran}

\PassOptionsToPackage{hyphens}{url}
\input{preamble.tex}
\usepackage{dsfont}
\usepackage{xr}

\newcommand{\CH}{\mathbb{C}^n\otimes\mathcal{H}}

\newacronym{GSO}{GSO}{graph shift operator}
\newacronym{WSS}{WSS}{wide-sense stationary}
\newacronym{JWSS}{JWSS}{joint wide-sense stationary}
\newacronym{VWSS}{VWSS}{vertex wide-sense stationary}
\newacronym{HWSS}{HWSS}{Hilbert space wide-sense stationary}
\newacronym{PSD}{PSD}{power spectral density}
\newacronym{JPSD}{JPSD}{joint power spectral density}
\newacronym{MSE}{MSE}{mean-square error}
\newacronym{DFT}{DFT}{discrete time Fourier transform}
\newacronym{BLUE}{BLUE}{best linear unbiased estimator}
\newacronym{AWGN}{AWGN}{additive white Gaussian noise}
\newacronym{LCE}{LCE}{linear conditional expectation}
\newacronym{ALCC}{ALCC}{average linear conditional operator}
\newacronym{GRP}{GRP}{graph random process}
\newacronym{MAP}{MAP}{maximum a posteriori}
\newacronym{GAE}{GAE}{graph auto-encoder}
\newacronym{GCN}{GCN}{graph convolutional neural network}
\newacronym{GNN}{GNN}{graph neural network}

\begin{document}
	
	\title{Wide-Sense Stationarity in Generalized Graph Signal Processing}
	\author{Xingchao~Jian, and Wee~Peng~Tay,~\IEEEmembership{Senior~Member,~IEEE}%
		\thanks{%
			This research is supported by the Singapore Ministry of Education Academic Research Fund Tier 2 grant MOE-T2EP20220-0002.  The authors are with the School of Electrical and Electronic Engineering, Nanyang Technological University, Singapore. 
			
			This paper has supplementary downloadable material available at \url{http://ieeexplore.ieee.org.}, provided by the author. The material includes a brief background of random elements on Hilbert spaces and proofs of background results listed in this paper. This material is 258KB in size.
		}%
	}
	
	
	
	\maketitle \thispagestyle{empty}
	
	
	\begin{abstract}
		We consider statistical graph signal processing (GSP) in a generalized framework where each vertex of a graph is associated with an element from a Hilbert space. This general model encompasses various signals such as the traditional scalar-valued graph signal, multichannel graph signal, and discrete- and continuous-time graph signals, allowing us to build a unified theory of graph random processes. We introduce the notion of joint wide-sense stationarity in this generalized GSP framework, which allows us to characterize a graph random process as a combination of uncorrelated oscillation modes across both the vertex and Hilbert space domains. We elucidate the relationship between the notions of wide-sense stationarity in different domains, and derive the Wiener filters for denoising and signal completion under this framework. Numerical experiments on both real and synthetic datasets demonstrate the utility of our generalized approach in achieving better estimation performance compared to traditional GSP or the time-vertex framework.
	\end{abstract}

	\begin{IEEEkeywords}
		Graph signal processing, Hilbert space, wide-sense stationarity, power spectral density.
	\end{IEEEkeywords}
	
	\section{Introduction}\label{sect:intro}
	
	\IEEEPARstart{I}{n} many applications such as those involving the brain network\cite{HuaLeaNic:J16}, sensor networks \cite{Jab:J17,TseLee:C19,GnaHumOud:C21} and image processing \cite{CheMagTanNg:J18,YagOzg:J20}, data may be naturally embedded within an underlying graph structure, which makes it beneficial for one to model such data as graph signals, i.e., a map from a vertex set to the Euclidean space of real values $\Real$ or complex values $\Complex$. Graph signal processing (GSP) theories and methods \cite{ShuNarFro:J13,OrtFroKov:J18} have been developed to leverage the graph structure to deal with the irregularities inherent in the graph domain, leading to new approaches to signal processing tasks for graph signals, including denoising\cite{YagOzg:J20}, sampling and recovery \cite{CheLeu:C16,LorBarBan:18,LinXieFenHu:C19,JiPraTay:C19,HarYamOnoTan:C21,GnaHumOud:C21}, and graph topology inference \cite{SarBarLor:J19,SegMarMatRib:J17,DonThaFroVan:J16}. GSP techniques like filters and wavelets also provide new insights, and improve efficiency and interpretability in machine learning \cite{DonThaTonBroFro:J20}. 
	
	Central to the theory of GSP is the concept of a \gls{GSO} and its resulting graph Fourier transform (GFT)\cite{Gir:C15,GirGonFle:J15,ShuNarFro:J13,SanMou:J13,JiTay:C21}, which is nothing but the ``frequency'' representation of a graph signal in a basis induced by the \gls{GSO} (typically, we assume a complete orthonormal eigenbasis consisting of eigenvectors of the \gls{GSO}). A \gls{GSO} and its corresponding GFT are graph-based counterparts of the translation operator and the discrete Fourier transform for a discrete time series. There are multiple choices of \gls{GSO}s and GFTs, which admit different properties and interpretations. For example, when the graph Laplacian is chosen as the \gls{GSO}, its eigenvectors represent signals of certain frequencies, i.e., each eigenvector is a representer of what a graph signal with a specific degree of ``smoothness'' along its edges looks like, with lower frequencies corresponding to smoother signals \cite{ShuNarFro:J13}.
	
	Analogies of signal processing concepts in the discrete time domain have been developed for graph domains in \cite{SanMou:J13}. For instance, similar to the time domain counterpart, a graph signal is bandlimited if its GFT support is within a proper subset of frequencies. Corresponding sampling theories on the graph domain have been developed \cite{CheLeu:C16,LorBarBan:18,LinXieFenHu:C19,JiPraTay:C19,HarYamOnoTan:C21}. Although the graph domain is irregular compared to the time domain, the concept of a \gls{WSS} graph signal or wide-sense stationarity (WSS) in the vertex domain can be defined analogously as that for a time domain signal \cite{Gir:C15,MarSegLeuRib:J17,PerVan:J17}. To be specific, the set of WSS signals is a family of signals with statistically uncorrelated spectral modes. The \gls{PSD} of a graph signal characterizes its energy distribution among different frequency components, and is also a crucial ingredient for constructing Wiener filters. 
	
	Traditional GSP deals with information in the graph or vertex domain rather than the time domain. To exploit information in both the graph and discrete-time domains, a time-vertex GSP theory\cite{PerLouGraVan:C17,LouPer:J19,GraLouPerRic:J18} was introduced. It includes the concepts of joint Fourier transform, joint WSS, joint PSD, and sampling strategies \cite{YuXieFenHu:C19}. 
	A further generalization to include vertex signals from a possibly infinite dimensional, separable Hilbert space was given in \cite{JiTay:J19}, which also developed the concepts of joint Fourier transform, filtering and sampling over the joint graph and Hilbert space domains. This framework encompasses a broad range of vertex signals including multichannel signals and continuous-time signals. In the sequel, we refer to this as the generalized GSP (GGSP) framework. Since we work only with separable Hilbert spaces, without loss of generality, we investigate only the Hilbert spaces $L^2(\Upsilon,\calG,\nu)$, where $\Upsilon$, $\calG$ and $\nu$ are, respectively, a suitably defined set, $\sigma$-algebra and measure, depending on the application of interest.
	
	The GGSP framework by \cite{JiTay:J19} does not consider randomness in the graph signal. In this paper, we fill this gap by introducing statistical elements to the GGSP framework, including the notion of \gls{WSS} graph signals over both the vertex and Hilbert space domains. In the accompanying supplementary material, we give a brief overview of the definitions and concepts of random elements in a Hilbert space and its moments. The main contributions of this paper are the following: 
	\begin{enumerate}
		\item We establish a probabilistic model for the GGSP framework by introducing the concept of a graph random process. We define the notion of joint WSS \gls{wrt} the GGSP shift operator. This framework includes the notion of graph or vertex WSS and joint WSS under the traditional GSP and time-vertex frameworks, respectively, as special cases. We show that joint WSS implies WSS in both the vertex and Hilbert space domains.
		\item We derive analytical forms for the Wiener filters for denoising and signal completion, which are the \gls{BLUE} for signal recovery under the joint WSS assumption. We also derive the theoretical \gls{MSE} of the Wiener filters as a criterion for sample set selection.
		\item We verify our proposed framework on real and synthetic datasets and demonstrate the utility of working with more general assumptions under our framework. In particular, the Hilbert space shift operator under our proposed framework can be learned from the data, instead of having to be fixed in advance.
	\end{enumerate}
	
	The rest of this paper is organized as follows. In \cref{sect:statmodel}, we model a random graph signal as a random element in a specific Hilbert space under the GGSP framework. In particular, we provide conditions for a stochastic process on a graph to be a random element. In \cref{sect:stationarity} we extend the joint WSS and \gls{JPSD} concepts from the time-vertex framework to the GGSP framework. Similar to \cite{PerLouGraVan:C17}, we investigate the relationship between joint WSS, vertex WSS and Hilbert space WSS. \cref{sect:wf} provides a Wiener filter design for denoising and signal completion. In \cref{sect:num_exp} we verify the effectiveness of our framework and techniques by numerical experiments on both synthetic and real-world datasets. Finally, we conclude in \cref{sect:conclusion}.
	
	\emph{Notations.} We write matrices and operators as boldface capital letters to distinguish them from scalars and vectors, which are written in plain lower cases. Random elements in a Hilbert space are denoted using plain capital letters. For any matrix $\bA$, we use $\bA\T$, $\bar{\bA}$, $\bA^*$ and $\vect(\bA)$ to denote its transpose, conjugate, conjugate transpose or adjoint and vectorization (columns stacked together to form a single-column vector), respectively.
	For a vector $a$, $\diag(a)$ is a diagonal matrix whose main diagonal is $a$. The matrix $\bI_n$ is the $n\times n$ identity matrix, $\bI_\calH$ is the identity operator on a Hilbert space $\calH$, and the $n$-dimensional all-ones column vector is $1_n$. We use the notation $\otimes$ to denote tensor product: $\calH_1\otimes\calH_2$ is the tensor product space of the vector spaces $\calH_1$ and $\calH_2$, which is spanned by the tensors $x\otimes y$ for $x\in\calH_1$ and $y\in\calH_2$. For operators $\bA_i$ on $\calH_i$, $i=1,2$, let $\bA_1\otimes\bA_2(x\otimes y) = \bA_1(x) \otimes \bA_2(y)$. Given two operators $\bC_1$ and $\bC_2$, $\bC_1\circ \bC_2$ (or abbreviated as $\bC_1\bC_2$) denotes their composition. The space of square integrable functions, $L^2(\Upsilon,\calG,\nu)$, where $\Upsilon$ is a set, $\calG$ a $\sigma$-algebra and $\nu$ a measure, is often abbreviated as $L^2(\Upsilon)$. The operator $\bPi_w$ is the projection operator on the subspace spanned by $w$.

	Common acronyms' full names and their first appearances in this paper are listed in \cref{tab:acro} for the reader's convenience.

	\begin{table}[!htbp]
		\centering
		\caption{List of acronyms}
		\label{tab:acro}
		
		\begin{tabular}{ p{0.16\columnwidth}|p{0.46\columnwidth}|p{0.16\columnwidth}  }
			Acronym& Full name &First appearance\\
			\hline
			GRP & Graph random process & \cref{def:random_graph_process}\\
			JWSS & Joint wide-sense stationary & \cref{def:JWSS} \\
			VWSS & Vertex wide-sense stationary & \cref{def:VWSS} \\
			HWSS & Hilbert space wide-sense stationary & \cref{def:HWSS}\\
			JPSD & Joint power spectral density & \cref{def:JWSS}\\
			LCE & Linear conditional expectation & \cref{subsect:noncausalWF}\\
			\hline
		\end{tabular}
		
	\end{table}

	\section{GGSP Statistical Model}\label{sect:statmodel}
	
	In this section, we present our model and assumptions. We define a \gls{GRP} in the GGSP framework as a random element and provide conditions under which it exists. Under the same conditions, the covariance operator of the \gls{GRP} is an integral operator. As a matrix kernel, its closed form enables us to interpret its trace via the generalized kernel method \cite{VitUmaVil:J13}.
	
	Let $\scG=(\scV,\scE)$ be an undirected (weighted) graph, where $\scV=\set{1,\ldots,n}$ is the vertex set, and $\scE\subset \set{(i,j)\in \scV\times \scV \given i<j}$ denotes the edge set. 
	Suppose that each vertex of $\scG$ is associated with an element from a separable Hilbert space $\calH$. A generalized graph signal \cite{JiTay:J19} has the form $x = (x_1,x_2,\ldots,x_n)$ where for each $v\in \scV$, we have $x_v\in\calH$. For example, if $\calH=L^2[a,b]$ the space of square integrable functions on a bounded interval $[a,b]$ (with the Borel $\sigma$-algebra and a given measure), then at each vertex $v$ of the graph $\scG$, we have associated with it a function $x_v = x_v(\cdot)\in L^2[a,b]$.
	
	It is shown in \cite{JiTay:J19} that the space of all generalized graph signals can be identified with the Hilbert space $\CH$ via the isomorphism
	\begin{align*}
		x \cong \sum_{v=1}^{n} \delta_v\otimes x_v,
	\end{align*}%
	where $\set{\delta_v \given v=1,\ldots,n}$ is the standard basis in $\Complex^n$. Let $\norm{}$ denote the norm of $\CH$.
	
	Suppose $\calH=L^2(\Upsilon)$ for some set $\Upsilon$. For $x \in \CH$ and each $t\in\Upsilon$, we write 
	\begin{align}\label{vector-form}
		x(t)=(x_1(t),x_2(t),\ldots,x_n(t))\T,
	\end{align}
	a vector-valued function.
	
	We next utilize the notion of a random element on a Hilbert space \cite{HsiEub:15,Bak:J73,Vak:87} to define a \gls{GRP}.
	
	\begin{Definition}\label{def:random_graph_process}
		Consider a probability space $(\Omega,\mathcal{F},\mu)$, where $\Omega$ is a sample space, $\calF$ a $\sigma$-algebra and $\mu$ a probability measure. A \emph{\gls{GRP}} $X$ is a measurable map (i.e., a random element) from $(\Omega,\mathcal{F})$ to $(\CH,\mathcal{B})$, where $\mathcal{B}$ is the Borel $\sigma$-algebra induced by the norm of $\CH$. 
	\end{Definition}
	
	\cref{def:random_graph_process} formally introduces the concept of random elements on the generalized graph signal space $\CH$, which was not studied in \cite{JiTay:J19}.
	
	To ensure the existence of a GRP $X$'s first and second moments, we only consider those $X$ satisfying the following assumption. $X$ induces a probability measure $\bbP$ on $(\CH,\calB)$ given by
	\begin{align*}
		\bbP(B)=\mu(X^{-1}(B)),~\forall B\in \calB.
	\end{align*}
	
	\begin{Assumption}\label{assumpt:mean_norm}
		\begin{align*}
			\bbE\norm{X}^2 &= \int_{\CH}  \norm{x}^2 \ud \bbP(x)<\infty.
		\end{align*}
	\end{Assumption} 
	
	Under \cref{assumpt:mean_norm}, $X$'s mean element $m_{X}$ and covariance operator $\bC_{X}$ are defined as follows: for all $u, v\in\calH$, 
	\begin{align}
		\angles{m_{X},u}
		&=\bbE[\angles{X,u}] =\int_{\CH} \angles{\xi,u} \ud\bbP(\xi), \label{def:mX}\\
		\angles{\bC_{X}u, v}
		&=\bbE[\overline{\angles{X-m_X,u}}\angles{X-m_X,v}]\nn
		&=\int_{\CH} \overline{\angles{\xi-m_X,u}}\angles{\xi-m_X,v} \ud\bbP(\xi). \label{def:CX}
	\end{align}
	By the Riesz representation theorem \cite[Theorem 3.7.7]{DebMik:00}, $m_{X}$ and $\bC_{X} (u)$ exist uniquely and are hence well-defined. To preclude the pathological cases (e.g., an uncorrelated complex random process' real and imaginary parts can be correlated), in this paper we only consider those random elements that are proper, i.e., satisfy the equivalent conditions in \cite[Theorem 1]{VakKan:J95}. It can be shown that covariance operators are bounded, trace-class and self-adjoint (cf.\ \cref{def:trace_class,def:selfadjoint_operator} in the supplementary). 
	
	The generalized concept of cross-covariance also appears as an operator. Suppose two random elements $X:\Omega\mapsto\calC_1$ and $Y:\Omega\mapsto\calC_2$, where $\calC_1,\calC_2$ are Hilbert spaces, induce a joint probability measure $\bbP_{XY}$ on $\calC_1\times\calC_2$. If $\bbE\norm{(X,Y)}^2<\infty$, the cross-covariance operator $\bC_{XY}:\calC_2\mapsto\calC_1$ is defined as 
	\begin{align}
		\angles{\bC_{XY}u,v}&=\bbE[\overline{\angles{Y-m_Y,u}}\angles{X-m_X,v}]\nn
		&=\int_{\calC_1\times\calC_2}\overline{\angles {\zeta-m_Y,u}}\angles{\xi-m_X,v} \ud\bbP_{XY}(\xi,\zeta), \label{def:CXY}
	\end{align}
	for all $u\in\calC_2$ and $v\in\calC_1$. We say that $X$ and $Y$ are uncorrelated if $\bC_{XY}=\bzero$. In this paper, we will only come across the case where $\calC_1=\calC_2=\CH$.
	
	Without loss of generality, we assume that any GRP $X$ of interest has mean element $m_X=0$ unless otherwise specified. Filters on $\CH$, as in \cite{JiTay:J19}, are defined as bounded linear operators on $\CH$.
	
	As an example, in \cref{def:random_graph_process}, $X$ can represent a multichannel signal. For instance, each vertex in a graph may correspond to a sensor station that records PM2.5, temperature and humidity levels, so that the recorded signal at each vertex is in $\calH=\Real^3$. If we take $X$ to be the random vector observations at all vertices, it satisfies \cref{def:random_graph_process} and is a \gls{GRP}.
	
	From the perspective of time-vertex analysis, it is also natural to suppose that each vertex signal can be a discrete- or continuous-time signal. We next investigate conditions under which the following stochastic process on $\scG$ can be modeled as a \gls{GRP} as in \cref{def:random_graph_process}: let $\calH=L^2(\Upsilon)$, where $\Upsilon$ is a set like $[a,b]$, and 
	\begin{align}
		\begin{aligned}\label{Xomegat}
			X:\Omega\times \Upsilon & \mapsto\Complex^n,\\
			(\omega,t)&\mapsto X(\omega,t).
		\end{aligned}
	\end{align}
	
	If $t$ is treated as an index so that $X(\cdot,t)$ is a function of $\omega$, and $X(\cdot,t)$ is measurable (\gls{wrt} $\omega$) for every $t$, then $\set{X(\cdot,t)\given t\in[a,b]}$ is a family of $n$-dimensional random vectors. We abbreviate $X(\cdot,t)$ as $X(t)$. 
	
	On the other hand, if $\omega$ is treated as an index so that $X(\omega,\cdot)$ is a function of $t$, $X(\omega,\cdot)$ can be interpreted as an $n$-dimensional trajectory under the outcome $\omega$. An immediate problem arises when no restrictions are imposed on $X$. In this case, the trajectory $\tilde{X}(\omega)=X(\omega,\cdot)$ can be arbitrarily irregular, and is not guaranteed to belong to $\Complex^n\otimes L^2(\Upsilon)$. Besides, even if the trajectories are restricted to $\Complex^n\otimes L^2(\Upsilon)$, without further constraints, $\tilde{X}(\omega)$ is not necessarily a measurable map from $\Omega$ to $\Complex^n\otimes L^2(\Upsilon)$, and hence may not fit the definition of a \gls{GRP} in \cref{def:random_graph_process}. To overcome these problems, we provide a sufficient condition in \cref{thm:contstime} to model $\tilde{X}$ as a random element in $\Complex^n\otimes L^2(\Upsilon)$. In \cref{thm:traceeq}, we derive the mean element and covariance operator. The proofs of \cref{thm:contstime} and \cref{thm:traceeq} are in \cref{sect:proof:thm1and2}.
	
	\begin{Theorem}\label{thm:contstime}
		Suppose $\calH=L^2(\Upsilon,\calG,\nu)$, where $\calG$ is a $\sigma$-algebra and $\nu$ a measure.
		The map $\tilde{X}(\omega)=X(\omega,\cdot)$ for $\omega\in\Omega$, where $X(\cdot,\cdot)$ is a map of the form \cref{Xomegat}, is a \gls{GRP} for $\CH$ if the following conditions hold: 
		\begin{enumerate}[(a)]
			\item\label[condition]{thm:contstime:jm} $X(\omega,t)$ is jointly measurable on $\Omega\times \Upsilon$ \gls{wrt} the product measure $\mu\times\nu$. 
			\item\label[condition]{thm:contstime:L2} $X(\omega,\cdot)\in \CH$ for every $\omega\in\Omega$. 
			\item\label[condition]{thm:contstime:mc} For every $s,t \in \Upsilon$, the pointwise mean $m_X(t):=\bbE[X(t)]\in\bbC^n$ and cross-covariance $\bK_X(s,t):=\cov(X(s),X(t))\in\bbC^{n\times n}$ are well-defined. Every entry of $\bK_X(s,t)$ as a function of $(s,t)\in\Upsilon\times\Upsilon$ belongs to $L^1(\Upsilon\times \Upsilon)$.
		\end{enumerate}
	\end{Theorem}
	\cref{thm:contstime} encompasses a broad range of signals that can be modeled as GRPs on $\CH$. For instance, let $\Upsilon=\Real_+$, $\calG$ the Borel $\sigma$-algebra and $\nu$ the Lebesgue measure. Let $\set{X(\omega,t)\given t\geq0}$ be a family of random variables indexed by $t$. If $X(\omega,\cdot)$ is continuous and \cref{thm:contstime:mc} in \cref{thm:contstime} is met, then all conditions in \cref{thm:contstime} are satisfied (see \cite[Proposition 1.13]{KarShr:98}). In another example, if $\Upsilon=\{1,2,\ldots,d\}$ (i.e., $\calH=\Real^d$), the conditions in \cref{thm:contstime} are met as long as $X(t)$ is a random vector with finite second moments for each $t=1,\ldots,d$. Hereafter where there is no confusion, we will not distinguish the maps $\tilde{X}$ and $X$, and simply write $X$ for both of them.
	
	We next show that the mean element $m_X$ and covariance operator $\bC_X$ of $X$ in \cref{def:mX,def:CX} follow from \cref{thm:contstime}. In particular, $\bC_X$ is an integral operator with kernel $\bK_X$.
	
	\begin{Theorem}\label{thm:traceeq}
		Suppose the conditions in \cref{thm:contstime} hold. Then, the mean element of $X$ as defined in \cref{def:mX} coincides with $m_X(t)$ in \cref{thm:contstime}, and the covariance operator of $X$ as defined in \cref{def:CX} is given by 
		\begin{align*}
			(\bC_Xf)(s)=\int_{\Upsilon} \bK_X(s,t)f(t) \ud \nu(t)
		\end{align*} 
		for all $f\in\CH$, where $f(t)$ in the integral is in the form \cref{vector-form}.
		Furthermore, suppose $\Upsilon$ is a separable metric space and $\nu$ is a finite measure on the Borel $\sigma$-algebra $\calB(\Upsilon)$ of $\Upsilon$. If $\bK_X(s,t)$ is continuous in $(s,t)$, the operator trace of $\bC_X$ agrees with the integral of its trace, i.e.,
		\begin{align*}
			\tr(\bC_X)=\int_{\Upsilon} \tr(\bK_X(t,t)) \ud\nu(t).
		\end{align*}
	\end{Theorem}
	
	From \cref{prop:tr=E} in the supplementary, the deviation of a random element from its mean can be measured by $\tr(\bC_X)$. To interpret \cref{thm:traceeq}, suppose $\Upsilon$ represents time domain. Then we have shown that this deviation amounts to performing integration of the pointwise variance on the time domain first, and then summing them up over all vertices, which also naturally measures the overall deviation.
	
	\section{Generalized Joint Wide-sense Stationarity}\label{sect:stationarity}
	
	In this section, we develop the notion of \gls{WSS} \gls{wrt} a shift operator for a \gls{GRP}. We define \gls{WSS} in different domains and study their relationships.
	
	\subsection{Joint WSS}\label{subsect:JWSS}
	
	Before we define the concept of joint \gls{WSS} for a \gls{GRP}, we briefly review the analogous concept of \gls{WSS} for a time domain scalar-valued stochastic process $x=(x_1,\ldots,x_m)$. This stochastic process $x$ is said to be \gls{WSS} if $\cov(x_i,x_j)$ only depends on $j-i$. Let $\bA_T$ be the one-step shift right operator, with wrapping to the front. The eigendecomposition of $\bA_T$ yields the \gls{DFT} matrix with columns being its eigenvectors. Then \gls{WSS} can be equivalently defined as $x$'s covariance matrix being diagonalizable by the DFT matrix.
	
	In the same spirit as the scalar case and noting that $\bA_T$ is the adjacency matrix of a directed cyclic graph, the analogous concept of graph wide-sense stationarity (GWSS)\cite{Gir:C15,MarSegLeuRib:J17,PerVan:J17} based on the \gls{GSO} \cite{SanMou:J13,GirGonFle:J15} have been proposed. Due to the intrinsic irregularity of the graph domain, there are multiple ways to define a \gls{GSO}. Widely-used choices include the graph adjacency matrix\cite{SanMou:J13}, Laplacian matrix\cite{ShuNarFro:J13} and an isometric \gls{GSO} design \cite{GirGonFle:J15}. 
	For GWSS, \cite{MarSegLeuRib:J17} provided equivalent definitions analogous to the aforementioned statements for \gls{WSS} of time domain scalar-valued signals. In \cite{PerVan:J17}, GWSS is defined by localization of a graph kernel.
	
	In this paper, to motivate a reasonable shift operator on $\CH$, we first investigate the product graph model. This model assumes the signal on each vertex to be a graph signal, hence is a special case where $\calH$ is a space of graph signals. This model allows for parallelized and vectorized implementations, and reduces the computational complexity for filters \cite{SanMou:J14}, and is thus an important model in practice. For a (weighted) graph $\scG$, we let $\bL_\scG$, $\bW_\scG$, $\bD_\scG$ be its graph Laplacian, adjacency matrix and (diagonal) degree matrix, respectively.
	
	\begin{Example}[Product graph]\label{exp:productgraph}
		Suppose each vertex of the graph $\scG=(\scV,\scE)$ observes a graph signal, which is defined on yet another graph $\scH=(\scV',\scE')$. This can be interpreted in the traditional GSP framework as signals on the vertices of a product graph $\scF$. There are multiple ways to construct the product graph $\scF$, including the tensor product graph $\scG\otimes \scH$ and the Cartesian product graph $\scG\times \scH$. In the case where $\scF=\scG\otimes\scH$, the adjacency matrix and graph Laplacian of $\scF$ can be written respectively as \cite{HamImrKla:11,BarBapPat:J15}:
		\begin{align}
			\bW_{\scG\otimes\scH}&=\bW_{\scG}\otimes \bW_{\scH}, \\
			\bL_{\scG\otimes\scH}&= \bD_{\scG}\otimes\bL_{\scH}+\bL_{\scG}\otimes\bD_{\scH}-\bL_{\scG}\otimes\bL_{\scH}.
		\end{align}
		In the case where $\scF=\scG\times\scH$, they become
		\begin{align}
			\bW_{\scG\times\scH}&=\bW_{\scG}\otimes\bI_{|\scV'|}+\bI_{|\scV|}\otimes\bW_{\scH}, \label{cprodW}\\
			\bL_{\scG\times\scH}&=\bL_{\scG}\otimes\bI_{|\scV'|}+\bI_{|\scV|}\otimes \bL_{\scH}. \label{cprodL}
		\end{align}
	\end{Example}
	
	Suppose we are given self-adjoint and compact operators $\bA_{\scG}$ on $\scG$ and $\bA_\calH$ on $\calH$. (From the Hilbert-Schmidt theorem on spectral decomposition, the choice of a self-adjoint and compact operator leads to an orthonormal eigenbasis consisting of eigenvectors of the chosen operator, which then allows us to define the Fourier transform. For finite dimensional spaces, compactness trivially holds.) Taking motivations from \cref{exp:productgraph}, a shift operator on $\CH$ can be chosen as either $\bA_{\scG}\otimes \bA_\calH$ or $\bA_{\scG}\otimes \bI_\calH+\bI_n\otimes \bA_\calH$. The operator $\bA_{\scG}\otimes \bA_\calH$ is the preferred over $\bA_{\scG}\otimes \bI_\calH+\bI_n\otimes \bA_\calH$ because the former is compact and self-adjoint, while the latter is not for infinite dimensional $\calH$. However, the eigenvectors of $\bA_{\scG}\otimes \bA_\calH$ are also the eigenvectors of $\bA_{\scG}\otimes \bI_\calH+\bI_n\otimes \bA_\calH$, hence the Fourier transform induced by both operators are the same. In the rest of this paper, we adopt $\bA_{\scG}\otimes \bA_\calH$ as the shift operator for $\CH$.
	
	Suppose $\bA_{\scG}$ has eigenvalues $\lambda_1\leq\lambda_2\ldots\leq\lambda_n$ with corresponding eigenvectors $\{\phi_k\}_{k=1}^{n}$, and $\bA_\calH$ has eigenvalues $\set{\nu_\tau}_{\tau=1}^{\infty}$ with corresponding eigenvectors $\{\psi_{\tau}\}_{\tau=1}^{\infty}$. Then, $\bA_{\scG}\otimes \bA_\calH$ has eigenvalues $\set{\lambda_k\nu_{\tau}\given k=1,\dots,n,\,\tau\geq1}$ and eigenvectors $\set{\phi_k\otimes\psi_{\tau} \given k=1,\dots,n,\,\tau\geq1}$.
	
	The joint Fourier transform of a signal $x \in \CH$ can be defined as its inner product with an eigenvector of the shift operator \cite{JiTay:J19}: for $k=1,\ldots,n$ and $\tau\geq1$,
	\begin{align}\label{JFT}
		\calF_x(\phi_k\otimes\psi_{\tau})=\angles*{x, \phi_k\otimes\psi_{\tau}}.
	\end{align}
	To simplify the exposition and to obtain a unique orthonormal eigenbasis (up to multiples of $\pm1$), similar to most of the GSP literature \cite{SanMou:J13,MarSegLeuRib:J17}, we make the following assumption throughout this paper. 
	\begin{Assumption}\label{assumpt:single-multiplicity}
		The geometric multiplicity of each eigenvalue of $\bA_{\scG}\otimes\bA_\calH$ is one.
	\end{Assumption}
	
	The convolution filter with coefficients $\set{g_{k,\tau}\given k=1,\ldots,n,\,\tau\geq1}$ is defined as the pointwise multiplication operator in the frequency domain:
	\begin{align}
		\bG (x) 
		&=\sum_{k,\tau} g_{k,\tau} \calF_x(\phi_k\otimes\psi_{\tau})\phi_k\otimes\psi_{\tau}\nn
		&=\sum_{k,\tau}g_{k,\tau}\bPi_{\phi_k\otimes\psi_{\tau}}(x), \label{convfilter}
	\end{align}
	where we recall that $\bPi_w$ is the projection operator onto the subspace spanned by $w$.
	
	\begin{Definition}\label{def:JWSS}
		A \gls{GRP} $X$ is \gls{JWSS} if 
		\begin{align}\label{eq:JWSS}
			\bC_X\circ (\bA_{\scG}\otimes \bA_\calH) = (\bA_{\scG}\otimes \bA_\calH)\circ\bC_X.
		\end{align}
		In other words, from \cref{assumpt:single-multiplicity} and \cite[Chapter 4, Exercise 35 (a)]{SteSha:05}, $\bC_X$ and the shift operator $\bA_{\scG}\otimes\bA_\calH$ have the same complete orthonormal set of eigenvectors with 
		\begin{align}\label{PSD-CX}
			\bC_X=\sum_{k,\tau} p_X(k,\tau)\bPi_{\phi_k\otimes\psi_{\tau}},
		\end{align}
		where $\set{p_X(k,\tau) \given k=1,\ldots,n,\,\tau\geq1}$ are the eigenvalues of $\bC_X$, also known as the \emph{\gls{JPSD}} of $X$. If $\set{(k,\tau)\given p_X(k,\tau)>0}$ is a finite set, $X$ is said to be \emph{bandlimited}.
	\end{Definition}
	
	In the following, we give two examples to illustrate \cref{def:JWSS}.
	
	\begin{Example}[Time-vertex model]\label{exp:time-vertex}
		Suppose each vertex of a graph $\scG$ observes a time series with $T$ discrete time steps. Then, the signal can be represented as a graph signal whose underlying graph $\scF$ is the Cartesian product graph of $\scG$ and a directed cyclic graph $\scH$ with $T$ vertices \cite{GraLouPerRic:J18}. Under this framework, we choose as the \gls{GSO} the Laplacian matrix $\bL_\scF$ of $\scF$. In \cite{LouPer:J19}, a time-vertex WSS signal is defined as WSS on the graph $\scF$ \gls{wrt} $\bL_{\scF}$.
		
		From \cref{cprodL}, the product graph Laplacian is $\bL_{\scF}=\bL_\scG\otimes\bI_T+\bI_n\otimes\bL_{\scH}$, which is different from the shift operator used in \cref{eq:JWSS}. Furthermore, $\bL_{\scH}$ is not a self-adjoint operator on $\Complex^T$. To be consistent with \cref{def:JWSS}, let $\widetilde{\scH}$ denote the undirected cyclic graph with $T$ vertices. Now we take $\bL_\scG\otimes\bL_{\widetilde{\scH}}$ as the shift operator. We first observe that $\bL_{\widetilde{\scH}}$ is self-adjoint and has the DFT matrix columns as eigenvectors. Therefore, the Fourier transform induced by $\bL_\scG\otimes\bL_{\widetilde{\scH}}$ is the same as that induced by $\bL_{\scF}$. Thus, a time-vertex WSS signal as defined in \cite{LouPer:J19} fits \cref{def:JWSS}.
	\end{Example}
	
	The time-vertex model is a special case in which the signal on each vertex belongs to a Euclidean space. In the next example, we will see that in a more general setting, the shift operator in the joint domain can also be defined in a meaningful way.
	
	\begin{Example}[Euclidean-vertex model]\label{exp:Euclidean-vertex}
		Suppose each vertex $v$ observes a $d$-dimensional random vector $x_v$ having identical distribution over vertices $v\in \scV$. The signal can be written as a matrix $\bX=(x_1,\ldots,x_n)\in\bbC^{d\times n}$, where the $v$-th column is the observation by vertex $v$ while the $i$-th row represents the $i$-th measurement or feature among all vertices. We choose the graph shift operator $\bA_{\scG}=\bL_{\scG}$, and $\bA_\calH$ as the covariance matrix $\bC_\calH=\bbE[x_vx_v^*]$ of $x_v$. In the context of \cref{def:JWSS}, we let the shift operator be $\bL_{\scG}\otimes\bC_\calH$.
		
		In this case, we note that the shift operator $\bL_{\scG}\otimes\bC_\calH$ has a physical meaning. Suppose $\bL_{\scG}=\bPhi\bLambda_{\scG}\bPhi\T$ and $\bC_\calH=\bPsi\bLambda_\calH\bPsi^*$ are eigendecompositions. The joint Fourier transform on the signal $\bX$ can be written as 
		\begin{align}\label{FcalX}
			\calF_\bX=\bPsi^*\bX\bPhi,
		\end{align}
		which in vectorized form $\vect(\calF_\bX)=\bm{\Phi}\T\otimes\bm{\Psi}^*\vect(\bX)$ corresponds to \cref{JFT}. The transform \cref{FcalX} can be interpreted as a two-step Fourier transform: first we multiply $\bPsi^*$ on $\bX$ on the left-hand side, so that each graph signal $x_v$ is transformed into its principal component scores. Next, we multiply $\bPhi$ on $\bPsi^*\bX$ on the right-hand side to transform each row into the graph frequency domain. 
		
		If $\bX$ is \gls{JWSS} \gls{wrt} $\bL_{\scG}\otimes\bC_\calH$, the covariance $\bC_\bX = \bbE[\vect(\bX)\vect(\bX)^*]$ of $\bX$ have eigenvectors given by the eigenbasis $\bPhi\otimes\bPsi$. From \cref{prop:PCA} in the supplementary, the entries of $\calF_\bX$ are uncorrelated random variables. 
		This implies that $\bX$ can be decomposed into different oscillation modes of its vertex signals' principal component scores along the edges of $\scG$. 
		In practice, $\bC_\calH$ can be estimated by the sample covariance of the vertex signal observations. Therefore, the shift operator and Fourier transform are deduced from both the graph topology and training data. Compared to the traditional GSP and time-vertex framework that construct the principal axes from the domain structures, this framework is closer to a data-driven approach.
		
		Although we have assumed that every vertex observes a $d$-dimensional random vector, our model is not equivalent to the product graph model, because $\bL_{\scG}\otimes\bC_\calH$ does not correspond to any kind of graph product in \cref{exp:productgraph}. However, one may adopt the shift operator $\bW_{\scG}\otimes\bC_\calH$ to obtain a tensor product graph model. In this case, $\bC_\calH$ is treated as the adjacency matrix of a weighted graph with $d$ vertices. But in this case, this model's physical meaning is unclear compared to the construction in this example. 
		
		Finally, we note that this model can be generalized such that the vertex signals $x_v$, $v\in \scV$, are not identically distributed. In fact, we only need to assume that every $x_v$ has the same set of principal axes, and regard $\bL_{\scG}\otimes\bC_{x_1}$ as the shift operator. It can be shown that this yields the same Fourier transform and physical meaning as discussed above.
	\end{Example}
	
	An example with infinite dimensional $\calH=L^2[-\pi,\pi]$ is presented in \cref{subsect:WSSdiffdom}.
	In the numerical experiments in \cref{sect:num_exp}, we employ the models in \cref{exp:time-vertex} and \cref{exp:Euclidean-vertex} to compare the effectiveness of their corresponding Wiener filters, which are discussed in \cref{sect:wf}.
	
	\subsection{WSS in different domains}\label{subsect:WSSdiffdom}
	In the time-vertex framework of \cite{LouPer:J19,PerLouGraVan:C17}, stationarity in the time domain (TWSS), vertex domain and joint domain are related in the sense that \gls{JWSS} implies stationarity in both the time and vertex domains. In  this subsection, the corresponding concepts and relations are generalized to the GGSP framework.
	
	\begin{Definition}\label{def:VWSS}
		Given a \gls{GRP} $X$ satisfying the conditions in \cref{thm:contstime}, we say that $X$ is \gls{VWSS} if
		\begin{align*}
			\bK_X(t,t)\bA_{\scG}=\bA_{\scG}\bK_X(t,t)
		\end{align*}
		$\nu$-almost everywhere (a.e.).
	\end{Definition}
	
	Note that \cref{def:VWSS} implicitly requires $\calH=L^2(\Upsilon)$ (an assumption in \cref{thm:contstime}) so that $\bK_X(t,t)$ can be defined. This does not result in loss of generality since we work only with separable Hilbert spaces. \Cref{def:VWSS} requires each measurement $X(t)$ at each ``time'' $t\in\Upsilon$ to be WSS as a vector-valued graph signal. For example, if $\Upsilon$ represents the time domain, this definition implies that the graph signal observed at every time instance is \gls{VWSS} as a random vector.
	
	\begin{Definition}\label{def:HWSS}
		A \gls{GRP} $X$ is said to be \gls{HWSS} if
		\begin{align}\label{eq:HWSS}
			\bdelta_m^\calH \bC_X\bdelta_m^\calH\circ (\bI_n\otimes \bA_\calH) = (\bI_n\otimes \bA_\calH)\circ\bdelta_m^\calH \bC_X\bdelta_m^\calH
		\end{align}
		for all $m=1,\ldots,n$. Here, $\bdelta_m^\calH:=\diag(\delta_m)\otimes \bI_\calH$ where $\delta_m$ is the $m$-th standard basis vector in $\bbC^n$ consisting of all zeros except a one at the $m$-th entry.
	\end{Definition}
	
	In \cref{def:HWSS}, the operator $\bdelta_m^\calH$ keeps the vertex signal at vertex $m$ unchanged while nullifying the other vertex signals. Suppose $X$ satisfies all conditions in \cref{thm:contstime}. For $y(\cdot)=(y_1(\cdot),\ldots,y_n(\cdot))\T\in \Complex^n\otimes L^2(\Upsilon)$, the left-hand side of \cref{eq:HWSS} can be computed step by step as follows:
	\begin{align}
		&\bI_n\otimes \bA_\calH (y)(t) =(\bA_\calH (y_1)(t),\ldots,\bA_\calH (y_n)(t))\T, \\
		&\bdelta_m^\calH \bC_X\bdelta_m^\calH\circ (\bI_n\otimes \bA_\calH(y))(s) \nn
		& =\int_{\Upsilon} \bK_X(s,t)_{m,m} \bA_\calH (y_m)(t)\,\ud \nu(t) \cdot\delta_m, \label{HWSS-LHS}
	\end{align}
	where $\bK_X(s,t)_{m,n}$ is the $(m,n)$-th entry of $\bK_X(s,t)\in\bbC^{n\times n}$.
	Similarly, the right-hand side of \cref{eq:HWSS} can be derived as 
	\begin{align}
		& (\bI_n\otimes \bA_\calH)\circ\bdelta_m^\calH \bC_X\bdelta_m^\calH (y)(s) \nn
		&=\bA_\calH \parens*{\int_{\Upsilon}\bK_X(\cdot,t)_{m,m}y_m(t)\ud \nu(t)}(s) \cdot\delta_m. \label{HWSS-RHS}
	\end{align}
	
	Let $\bK_m$ be the integral operator with kernel $\bK_X(s,t)_{m,m}$ in the right-hand side of \cref{HWSS-RHS}. Comparing \cref{HWSS-LHS} and \cref{HWSS-RHS}, for $X$ to be \gls{HWSS}, the operator $\bA_\calH$ commutes with the integral operator $\bK_m$. 
	
	Consider the example where $\Upsilon=[-\pi,\pi]$ with $\bA_\calH$ being a convolution operator,
	\begin{align*}
		\bA_\calH(y)=\sum_{k=-\infty}^{\infty} \nu_{|k|}\bPi_{\psi_{k}},
	\end{align*}
	where $\{\nu_k\}_{k=0}^{\infty}$ is a positive sequence converging to 0, and $\set{\psi_{k} \given k\in\mathbb{Z}}$ denotes the standard Fourier basis $\set{e^{\iu kt}/\sqrt{2\pi} \given k\in\mathbb{Z}}$, where $\iu=\sqrt{-1}$. Suppose $\bK_X(s,t)$ is continuous on $\Upsilon\times \Upsilon$. In this case, both $\bA_\calH$ and $\bK_m$ are compact and self-adjoint operators on $L^2(\Upsilon)$. If $\bA_\calH$ commutes with $\bK_m$, then since the standard Fourier basis $\{\psi_{k}\}_{k=0}^{\infty}$ are eigenvectors of $\bA_\calH$, they are also the eigenvectors of $\bK_m$. Since
	\begin{align*}
		\sum_{i,j}^n c_i\bar{c}_j\bK_X(s_i,s_j)_{m,m}=\var\parens*{\sum_i^n c_iX_m(s_i)} \geq 0
	\end{align*}
	for arbitrary $\set{c_i}_{i=1}^n\subset\Complex$ and $\set{s_i}_{i=1}^n\subset \Upsilon$, $\bK_X(s,t)_{m,m}$ is a positive definite symmetric kernel on $\Upsilon$. Utilizing Mercer's theorem we obtain that 
	\begin{align*}
		\bK_X(s,t)_{m,m}=\frac{1}{2\pi}\sum_{k=-\infty}^{\infty} \sigma_k e^{\iu ks}e^{-\iu kt}=\frac{1}{2\pi}\sum_{k=-\infty}^{\infty} \sigma_k e^{\iu k(s-t)},
	\end{align*}
	where $\set{\sigma_k \given k\in\mathbb{Z}}$ are $\bK_m$'s eigenvalues associated with $\set{\psi_{k} \given k\in\mathbb{Z}}$. This infinite sum uniformly converges on $\Upsilon\times \Upsilon$, indicating that $\bK_X(s,t)_{m,m}$ is actually a univariate function of $s-t$ for each $m$. This result agrees with the classical definition of stationarity in the time domain.
	
	\cref{def:VWSS,def:HWSS} can be regarded as traditional WSS definitions embedded in the GGSP framework. In the following theorem we show that \gls{JWSS} implies WSS in both the vertex and Hilbert space domains.
	
	\begin{Theorem}\label{thm:JWSS and HWSS}
		A \gls{JWSS} \gls{GRP} $X$ that satisfies the conditions in \cref{thm:contstime} is both \gls{VWSS} and \gls{HWSS}.
	\end{Theorem}
	\begin{IEEEproof}
		We first show that $X$ is \gls{VWSS}. Since every $\bPi_{\phi_k\otimes\psi_{\tau}}$ commutes with $\bA_{\scG}\otimes \bI_\calH$, from the fact that $X$ is \gls{JWSS}, (cf.\ \cref{def:JWSS}), we obtain that $\bC_X$ commutes with $\bA_{\scG}\otimes \bI_\calH$. To be specific, for any $y\in\Complex^n\otimes L^2(\Upsilon)$, 
		\begin{align*}
			\bC_X\circ (\bA_{\scG}\otimes \bI_\calH)(y)(s)=\int_{\Upsilon} \bK_X(s,t)\bA_{\scG} y(t)\ud \nu(t),\\
			(\bA_{\scG}\otimes \bI_\calH)\circ \bC_X(y)(s)=\bA_{\scG}\int_{\Upsilon}\bK_X(s,t)y(t)\ud \nu(t),
		\end{align*}
		hence 
		\begin{align*}
			\int_{\Upsilon} (\bK_X(s,t)\bA_{\scG}-\bA_{\scG}\bK_X(s,t))y(t)\ud \nu(t)=\bzero.
		\end{align*}
		Therefore, $\bK_X(s,t)\bA_{\scG}=\bA_{\scG}\bK_X(s,t)$ $\nu$-a.e., indicating that $X$ is \gls{VWSS}. 
		
		We can show that $X$ is \gls{HWSS} by a similar approach. It suffices to notice from \cref{PSD-CX} that 
		\begin{align*}
			\bdelta_m^\calH \bC_X\bdelta_m^\calH=\sum_{k,\tau} p_X(k,\tau)\cdot\bdelta_m^\calH \bPi_{\phi_k\otimes\psi_{\tau}}\bdelta_m^\calH,
		\end{align*}
		and each $\bdelta_m^\calH \bPi_{\phi_k\otimes\psi_{\tau}}\bdelta_m^\calH$ commutes with $\bI_n\otimes \bA_\calH$.
	\end{IEEEproof}
	
	From the proof of \cref{thm:JWSS and HWSS}, we note that \gls{JWSS} is strictly stronger than \gls{VWSS}, since we actually show that \gls{JWSS} implies $\bK_X(s,t)\bA_{\scG}=\bA_{\scG}\bK_X(s,t)$, which is a stronger condition than the \gls{VWSS} condition of $\bK_X(t,t)\bA_{\scG}=\bA_{\scG}\bK_X(t,t)$. In fact, \gls{JWSS} is strictly stronger than both \gls{VWSS} and \gls{HWSS}. This can be seen from \cite{LouPer:J19}, which introduces MTWSS (multivariate time WSS) and M\gls{VWSS} (multivariate vertex WSS) so that \gls{JWSS} is equivalent to simultaneously satisfying these conditions. These two concepts require not only the covariance but also the cross-covariance in their respective domains to admit certain forms of eigendecomposition. Our definitions of \gls{VWSS} and \gls{HWSS} extend the concept of \gls{VWSS} and TWSS defined in \cite{PerLouGraVan:C17}, which only require the covariance matrices to satisfy the conditions in \cref{def:VWSS,def:HWSS}.
	
	\section{Wiener Filters}\label{sect:wf}
	
	In this section, we investigate the denoising and recovery problems in the GGSP framework. In traditional GSP, these problems are formulated as regularized regression problems, whose regularization terms depend on the PSD values of the signal and noise\cite{PerVan:J17}. This optimization framework is also adopted under the time-vertex framework \cite{PerLouGraVan:C17}. The Wiener filters of these frameworks are the \gls{BLUE} in their respective frameworks. 
	In the GGSP framework, the observed signal on each vertex may come from an infinite-dimensional Hilbert space. In the sequel, we see that the Wiener filter takes the same form as the aforementioned formulations.
	
	\subsection{Wiener filter for denoising}\label{subsect:noncausalWF}
	
	Consider the model
	\begin{align}\label{eq:denoimod}
		Y=X+\calE
	\end{align}
	where $X$ and $\calE$ denote the signal and noise, respectively. Suppose $X$ and $\calE$ are independent \gls{JWSS} \gls{GRP}s. From \cref{prop:covtrans} in the supplementary, $Y$ is then a \gls{JWSS} \gls{GRP} with its JPSD given by $\set{p_X(k,\tau)+p_\calE(k,\tau) \given k=1,\ldots,n,\,\tau\geq1}$. We further assume that $m_X=m_{\calE}=0$ for simplicity.
	
	Deriving the Wiener filter, in this case, amounts to deriving the \gls{BLUE} for $X$ given $Y$. In Hilbert space, this corresponds to \gls{LCE} (see \cref{sect:LCE} of the supplementary or \cite{KleSprSul:J20}). These results enable us to derive an explicit formula for the Wiener filter.          
	
	\begin{Theorem}\label{thm:noncausalWF}
		The Wiener filter $\bG$ corresponding to the model \cref{eq:denoimod} is a convolution filter of the form \cref{convfilter} with coefficients
		\begin{align}\label{eq:wienerco}
			g_{k,\tau}=\frac{p_X(k,\tau)}{p_X(k,\tau)+p_{\calE}(k,\tau)}.
		\end{align}
	\end{Theorem}
	\begin{IEEEproof}
		From \cref{prop:incompatible} in the supplementary, $\bG$ can be asymptotically approximated by a sequence of bounded finite-rank operators $\{\bG^{(m)}\}_{m\geq1}$ as follows: Let $\scV^{(m)}=\spn\set{\phi_k\otimes\psi_{\tau} \given k=1,\ldots,n,\,\tau=1,\ldots,m}$ and $Y^{(m)}=\bPi_{\scV^{(m)}}Y$, then for any $y\in\CH$,
		\begin{align}
			\bG^{(m)}(y) &=(\bC_{Y^{(m)}}^\dag \bC_{Y^{(m)}X})^*y, \label{eq:Gm}
		\end{align}
		where $^\dag$ denotes the Moore-Penrose pseudoinverse. Since $Y$ is \gls{JWSS}, $\bC_Y$ has the eigendecomposition
		\begin{align*}
			\bC_Y=\sum_{\tau=1}^{\infty}\sum_{k=1}^{n} (p_X(k,\tau)+p_{\calE}(k,\tau))\bPi_{\phi_k\otimes\psi_{\tau}},
		\end{align*}
		Using \cref{prop:covtrans} in the supplementary, the terms on the right-hand side of \cref{eq:Gm} can be written as 
		\begin{align*}
			\bC_{Y^{(m)}}^\dag &=\sum_{\tau=1}^{m}\sum_{k=1}^{n}(p_X(k,\tau)+p_{\calE}(k,\tau))^{-1} \bPi_{\phi_k\otimes\psi_{\tau}},\\
			\bC_{Y^{(m)}X}&=\bPi_{\scV^{(m)}} \bC_{YX}\\
			&=\bPi_{\scV^{(m)}} \bC_X\\
			&=\sum_{\tau=1}^{m}\sum_{k=1}^{n}p_X(k,\tau) \bPi_{\phi_k\otimes\psi_{\tau}}.
		\end{align*}
		By substituting these into \cref{eq:Gm}, we obtain
		\begin{align*}
			\bG^{(m)}(y)=\sum_{\tau=1}^{m}\sum_{k=1}^{n}\frac{p_X(k,\tau)}{p_X(k,\tau)+p_{\calE}(k,\tau)}\bPi_{\phi_k\otimes\psi_{\tau}}(y).
		\end{align*}
		From \cref{prop:incompatible} in the supplementary, $\norm{\bG^{(m)}(y)-\bG(y)} \to 0 $ as $m\rightarrow\infty$ for $\bbP_Y$-almost surely all $y\in\CH$. Therefore, the Wiener filter $\bG$ can be chosen to be
		\begin{align}\label{eq:wienerft}
			\bG(y) 
			&=\lim_{m\rightarrow\infty}\bG^{(m)}(y)\\
			&= \sum_{\tau=1}^{\infty}\sum_{k=1}^{n}\frac{p_X(k,\tau)}{p_X(k,\tau)+p_{\calE}(k,\tau)}\bPi_{\phi_k\otimes\psi_{\tau}}(y),
		\end{align}
		yielding the result in \cref{eq:wienerco}. 
		
		Before concluding the proof, we remark that \cref{eq:wienerft} may not converge with respect to the operator norm induced topology. However, this infinite sum is still well-defined in terms of the strong operator topology, i.e.,  $\lVert \bG^{(m)}(y)-\bG(y)\lVert\rightarrow 0$ for all $y\in\calH$. Finally, it is straightforward to see that $\bG$ is a bounded linear operator with $\norm{\bG} \leq1$.
	\end{IEEEproof}
	
	\subsection{Wiener filter for signal completion}\label{subsect:sigcomple}
	
	We now consider the case where only signals from a subspace $\calA\subset\CH$ are observable, i.e., 
	\begin{align}
		Y=\bPi_\calA(X+\calE),
	\end{align}
	where $\bPi_\calA$ denotes the projection operator onto $\calA$. 
	
	In this case it may not be possible to give an explicit formula for the Wiener filter. However, if we assume that the signal is bandlimited (cf.\ \cref{def:JWSS}), the noise can be assumed to be bandlimited as well without loss of generality. This is because we can apply the finite-rank projection operator on the observed signal, projecting it to the subspace where the original signal lies in. By doing this, the frequencies that do not involve $X$ are discarded. Under this assumption, there is an explicit characterization of the Wiener filter as follows.
	
	\begin{Theorem}\label{thm:casualWF}
		Suppose $X$ and $\calE$ are bandlimited. The Wiener filter $\bG$ for signal completion can be written as 
		\begin{align}\label{eq:casualWF}
			\bG(y)=((\bPi_\calA(\bC_X+\bC_{\calE})\bPi_\calA)^\dag\bPi_\calA \bC_X)^*y,
		\end{align}
		for $y\in\CH$. In particular, when $\calH$ is finite-dimensional, $X$ and $\calE$ are trivially bandlimited.
	\end{Theorem}
	\begin{IEEEproof}
		Since $X$ and $\calE$ are bandlimited, their covariance operators are finite-rank. Notice that $\bC_Y=\bPi_\calA(\bC_X+\bC_{\calE})\bPi_\calA$, hence $\bC_Y$'s range is also finite-rank, thus closed. From \cref{prop:cond_for_compat} in the supplementary, the formula for the compatible case (cf.\ \cref{prop:compatible} in the supplementary) can be applied to directly obtain the result in \cref{eq:casualWF}.
	\end{IEEEproof}
	
	When $X$ and $\calE$ are not bandlimited, the Wiener filter can be approximated by a sequence of operators as in the proof of \cref{thm:noncausalWF}. 
	
	\begin{Theorem}\label{thm:causalWFapprox}
		The Wiener filter $\bG$ for signal completion can be asymptotically approximated by
		\begin{align*}
			& \bG_m(y)\\
			&=((\bPi_{\scV^{(m)}}\bPi_\calA(\bC_X+\bC_{\calE})\bPi_\calA\bPi_{\scV^{(m)}})^\dag\bPi_{\scV^{(m)}}\bPi_\calA \bC_X)^*y,
		\end{align*}
		where $\scV^{(m)}=\spn\set{\phi_k\otimes\psi_{\tau} \given k=1,\ldots,n,\,\tau=1,\ldots,m}$, and $\norm{\bG_m(y)-\bG(y)}\to 0$ for almost surely all $y\in\CH$.
	\end{Theorem}
	\begin{IEEEproof}
		This theorem is a direct result of \cref{prop:incompatible} in the supplementary.
	\end{IEEEproof}
	
	In the specific case that $\calA$ is the subspace of signals that annihilate on a set of vertices (i.e., only signals on a subset of vertices $\scU$ is observable), the analytical form of the signal completion error can be computed. Let $\bPhi$ be the matrix whose columns are $\set{\phi_1,\ldots,\phi_n}$, the eigenvectors of $\bA_{\calG}$. To obtain the \gls{MSE} of $\bG$, we define the following matrices $\bm{\Lambda}_\tau$ and $\bm{\Gamma}_{\tau}$ whose $(k,l)$-th elements are respectively:
	\begin{align*}
		(\bm{\Lambda}_\tau)_{k,l}&=
		\begin{cases}
			p_X(k,\tau)+p_{\calE}(k,\tau), & \text{if } k=l,\\
			0, & \text{if } k\neq l,
		\end{cases}\\
		(\bm{\Gamma}_{\tau})_{k,l}&=
		\begin{cases}
			p_X(k,\tau)^2, & \text{if } k=l,\\
			0, & \text{if } k\neq l.
		\end{cases}
	\end{align*}
	Let $\bPhi_{\scU}$ be the submatrix of $\bPhi$ that contains the rows with index set $\scU$.
	
	\begin{Theorem}\label{thm:mse}
		Assume that there exists $\tau_0<\infty$ such that whenever $\tau<\tau_0$, $p_X(k,\tau)+p_\calE(k,\tau)>0$ for all $k=1,\dots n$; and whenever $\tau\geq\tau_0$, $p_X(k,\tau)+p_\calE(k,\tau)=0$ for all $k=1,\dots n$. Then the \gls{MSE} of the Wiener filter $\bG$ can be written as
		\begin{align}
			& \bbE\norm*{\bG(Y)-X}^2 \nn
			&=\sum_{\tau=1}^{\tau_0}\sum_{k=1}^{n} p_X(k,\tau)-\sum_{\tau=1}^{\tau_0}\tr((\bPhi_{\scU}\bm{\Lambda}_\tau\bPhi_{\scU}\T)^{-1}\bPhi_{\scU}\bm{\Gamma}_\tau\bPhi_{\scU}\T). \label{eq:theomse}
		\end{align}
	\end{Theorem}
	\begin{IEEEproof}
		The basic idea of the proof is to characterize the image and kernel space of $\bC_Y$ first (i.e. \cref{lem:C_Ystructure} in the supplementary), and then compute the MSE via \cref{prop:tr=E} in the supplementary. See \cref{sect:proof:thmmse} of the supplementary for details.
	\end{IEEEproof}
	
	\cref{thm:mse} indicates a method to measure the quality of a sampling strategy.
	When the statistical properties of the signal and noise are fixed, one can choose the sampling set $\scU$ that admits a large value of $\sum_{\tau=1}^{\tau_0}\tr((\bPhi_{\scU}\bm{\Lambda}_\tau\bPhi_{\scU}\T)^{-1}\bPhi_{\scU}\bGamma_\tau\bPhi_{\scU}\T)$ to obtain a small \gls{MSE}.

	\section{Numerical Experiments}\label{sect:num_exp}
	
	In this section, we verify the performance of our proposed \gls{GRP} framework on four datasets. We compare its performance with the time-vertex and traditional GSP frameworks, and demonstrate that with its more general assumptions, the \gls{GRP} can fit data better, thus achieving better performance. We also illustrate the optimality of the proposed Wiener filter and the criterion of sample set selection through the experiments.
	
	Each dataset used can be organized into samples $\set{\bX_i \given i=1,\dots,m_s}$, where $\bX_i\in\Real^{n\times d}$, $n$ is the number of vertices in a graph $\scG$, and $d$ is the dimension of each data feature vector. We assume that $\bX_i$ are the matrix form of \gls{iid} realizations of a \gls{JWSS} \gls{GRP} $X$.
	
	Each of the GSP frameworks has different statistical model assumptions, under which PSD estimation or covariance estimation from a training set of size $m_a$ samples is performed. Throughout, we let the graph shift operator $\bA_\scG$ be the graph Laplacian with eigenbasis $\set{\phi_k\given k=1,\ldots,n}$. In the following, we present the concrete implementation of the PSD estimator under each framework when $d<\infty$. The case $d=\infty$ is discussed in \cref{subsect:contrec}.
	\begin{enumerate}
		\item \gls{GRP}. 
		The datasets we test on contain vertex signals from finite-dimensional real spaces, which fit the Euclidean-vertex model in \cref{exp:Euclidean-vertex}. 
		To estimate the vertex signal covariance $\bC_\calH$, we use the sample covariance matrix $\widehat{\bC}_\calH\in\Real^{d\times d}$ of a set of training samples. Let the eigenbasis induced by $\widehat{\bC}_\calH$ be $\hbPsi=\set{\hat\psi_{\tau} \given \tau=1,\ldots,d}$. 
		
		The JPSD values of the \gls{GRP} $X$ is estimated by its empirical mean squared Fourier coefficients through 
		\begin{align}\label{JPSD-empirical}
			\hat{p}_X(k,\tau)= \ofrac{m_a} \sum_{i=1}^{m_a} |\phi_{k}\T\bX_i\bar{\psi}_{\tau}|^2,
		\end{align}
		for $k=1,\ldots,n$ and $\tau=1,\ldots,d$. The covariance $\bC_X$ is estimated by 
		\begin{align}\label{eq:exp:H-V_C_X_estimator}
			\widehat{\bC}_X=\sum_{k,\tau} \hat{p}_X(k,\tau) (\phi_k\otimes\hat\psi_\tau)(\phi_k\otimes\hat\psi_\tau)\T.
		\end{align}
		
		\item Time-vertex (TV) \cite{LouPer:J19}. The \gls{JWSS} model and sample JPSD estimator proposed in \cite{LouPer:J19} are adopted. Specifically, the JPSD and $\bC_X$'s estimators are the same as those for the \gls{GRP} model, with the exception that the eigenbasis $\hbPsi$ is replaced by the column vectors of the DFT matrix rather than learned from the sample set.
		
		\item Traditional GSP. Since the number of features $d>1$ at each vertex, to adopt the traditional GSP framework that assumes scalar-valued vertex signals, we process the $d$ features separately. To be specific, write each sample $\bX_i=(F_{i,1},\ldots,F_{i,d})$, where $F_{i,j}\in\Real^n$, $j=1,\ldots,d$, is the $j$-th column and contains the $j$-th feature of all vertices. We use the periodogram \cite{MarSegLeuRib:J17} to estimate $p_X(k,\tau)$, $k=1,\ldots,n$, $\tau=1,\ldots,d$:
		\begin{align*}
			\hat{p}_X(k,\tau)= \ofrac{m_a} \sum_{i=1}^{m_a} |\phi_k\T F_{i,\tau}|^2.
		\end{align*}
		The covariance estimator for feature $\tau=1,\ldots,d$ is constructed similarly as \cref{eq:exp:H-V_C_X_estimator}:
		\begin{align*}
			\widehat{\bC}_{X,\tau} =\sum_{k} \hat{p}_X(k,\tau) \phi_k\phi_k\T.
		\end{align*}
		A separate Wiener filter for each feature is then constructed.
		
	\end{enumerate}
	In the experiments, all Wiener filter forms are applied with full bandwidth for fair comparison unless otherwise specified.

	\subsection{Wiener filter for denoising}\label{subsec:exp:noncau_WF}
	
	We investigate the denoising performance of the Wiener filter on an epilepsy dataset \cite{KraKolKir:J08}, which is collected by monitoring a patient's brain signal.\footnote{https://math.bu.edu/people/kolaczyk/datasets.html} The dataset is collected from 76 electrodes, during ictal and pre-ictal periods. In our experiments, due to the assumption of stationarity, we only make use of the pre-ictal data, which contains 8 pre-ictal periods. Each period lasts for 10s, and we partition it into non-overlapping 125ms periods due to the intrinsic long-term instability of a brain signal. Since the sampling rate is 400Hz, this partition means that each signal sample $\bX_i\in\Real^{76\times50}$, where $i=1,\ldots, 640$. We use the samples from the first 4 pre-ictal periods as training and the rest of them for testing. 
	
	In order to embed a signal sample in a graph structure, we use a simpler but similar strategy as that in \cite{KraKolKir:J08} to determine the connections between electrodes. Specifically, $4.5$-$5.5$s of data in the first pre-ictal period is extracted. Then the correlation matrix of this sample set is computed. Assuming that a large absolute value of correlation indicates strong connection, we treat each node pair as connected with an edge if the absolute value of their correlation coefficient is larger than $0.75$. Otherwise, they are not connected. 
	
	\begin{figure}[!htb]
		\centering
		\includegraphics[width=\linewidth, trim=.5cm 0cm .5cm .5cm, clip]{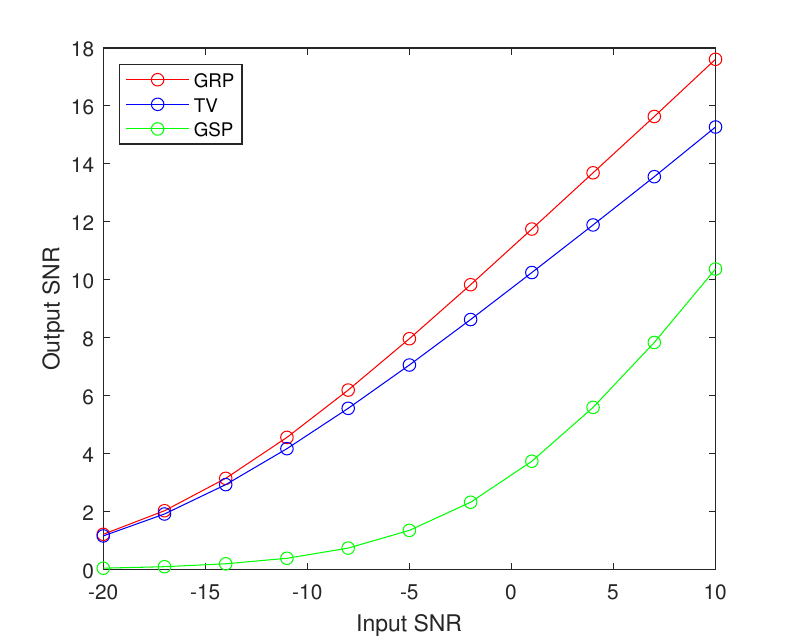}
		\caption{Denoising performance of Wiener filters under different frameworks. Experiments are repeated 20 times for each input SNR value.}
		\label{fig:WF}
	\end{figure}
	
	To compare the performance of different denoising strategies, the pre-ictal datasets are divided into training and test sets with the same size. \Gls{AWGN} with different energies is added to both of these sets to obtain different input SNRs. 
	Here, SNR in dB is defined as 
	\begin{align*}
		\text{SNR} = 10\log_{10} \frac{\bbE\norm{X}^2}{\bbE\norm{\widehat{X}-X}^2},
	\end{align*}
	where $X$ is the original signal. For input SNR, $\widehat{X}$ is the noisy version of $X$; for output SNR, $\widehat{X}$ denotes the estimate of $X$.
	
	
	By learning the signal spectrum from the training set, we aim to recover the signal from the noisy test set. The corresponding output SNR is taken as a measurement of performance. We compare the \gls{GRP} Wiener filter's performance with the time-vertex joint Wiener filter corresponding to the solution of the optimization framework proposed in \cite{LouPer:J19}, which has been shown to outperform both purely time-based and graph-based Wiener filters. In this experiment we also include the traditional GSP Wiener filter as a benchmark method. We have tested the purely time-based Wiener filter on this dataset, but the performance is much worse than the other methods, and is omitted here.
	
	From \cref{fig:WF}, we observe that the \gls{GRP} framework produces the highest output SNR compared to the other benchmark methods. This result indicates that the strategy of learning $\hbPsi$ from the training set provides a better fit than using the DFT basis.
	
\begin{figure}[!htb]
		\centering
		\includegraphics[width=\linewidth, trim=.5cm 0cm .5cm .5cm, clip]{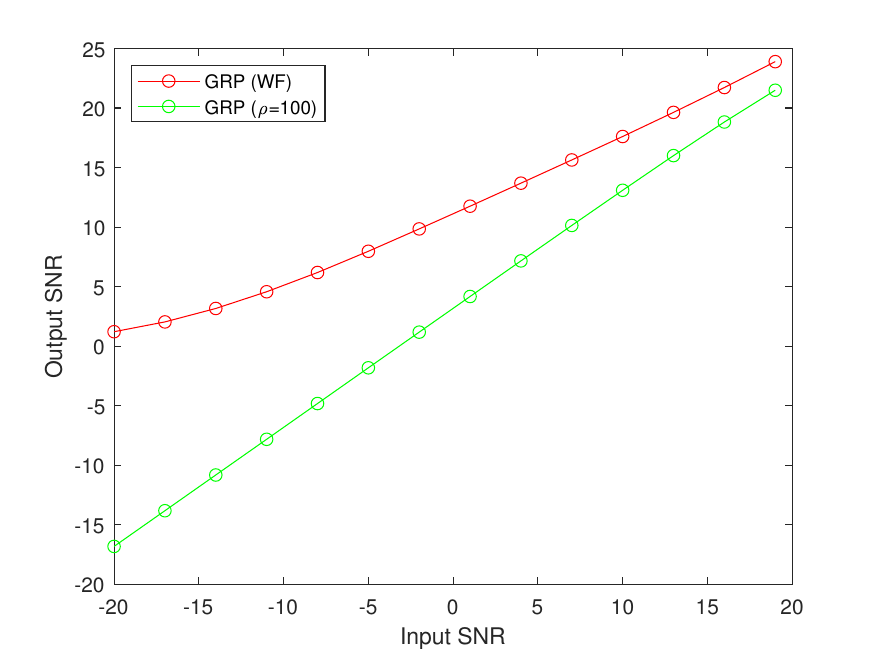}
		\caption{Denoising performance of different filter forms under GRP. Experiments are repeated 20 times for each input SNR value. ``GRP (WF)'' denotes the Wiener filter \cref{eq:wienerco}, and ``GRP ($\rho=100$)'' denotes \cref{GRPrho}.}
		\label{fig:GRPfilters}
	\end{figure}

	We want to test the denoising performance of the GRP Wiener filter versus other parameterized filters. A common strategy (see e.g., Example 2 in \cite{ShuNarFro:J13}) in traditional GSP is to solve the following problem: 
	\begin{align*}
		\hat{X}(Y)=\argmin_{X\in\Real^n} \norm{X-Y}^2+\rho X\T\bL X,
	\end{align*}
	where $\rho$ controls the smoothness of the recovered signal.
	On the other hand, the \gls{MAP} estimator for Gaussian random vectors in $\Real^d$ under the model \cref{eq:denoimod} is obtained as follows:
	\begin{align*}
		\hat{X}(Y)=\argmin_{X\in\Real^d} \norm{X-Y}^2+\sigma^2 X\T\bC_X^\dag X,
	\end{align*}
	where $\sigma^2$ denotes the variance of the noise. 
	Combining the above optimization problems, we obtain the following problem under the \gls{GRP} model:
	\begin{align}
		\hat{X}(Y)=&\argmin_{X\in\Real^{n\times d}} \norm{\vect(X)-\vect(Y)}^2\nn
		&+ \rho\vect(X)\T(\bL\otimes\bC_\calH^\dag) \vect(X), \label{GRPrho}
	\end{align}
	which is equivalent to applying the convolution filter with coefficients
	\begin{align*}
		\tilde{g}_{k,\tau} = \frac{\nu_\tau}{\nu_\tau+\rho\lambda_k},
	\end{align*}
	where $\set{\nu_\tau}$ and $\set{\lambda_k}$ are the eigenvalues of $\bC_\calH$ and $\bL$, respectively. We select the optimal parameter $\rho$ with the best performance, and compare it with the Wiener filter $\bG$ in \cref{thm:noncausalWF} under the same setting as \cref{fig:WF}. From \cref{fig:GRPfilters}, we see that the Wiener filter $\bG$ derived in \cref{eq:wienerco} outperforms the parameterized filter in \cref{GRPrho} as it is \gls{BLUE}.
	
	\subsection{Wiener filter for signal completion}\label{subsec:exp:causal_WF}
	
	We next evaluate the Wiener filter for signal completion on the Krakow air quality dataset, which contains air quality data from a sensor network in Krakow, Poland.\footnote{\url{https://www.kaggle.com/datascienceairly/air-quality-data-from-extensive-network-of-sensors}} The network consists of 56 sensors deployed across the city and each taking measurements on an hourly basis throughout the year 2017. We regard each sensor as a vertex in a graph. Each sensor records six measurements, namely the PM1, PM2.5, PM10, temperature, air pressure and humidity values. In this experiment we only consider pollution-related features, i.e., PM1, PM2.5 and PM10. Thus, this dataset can be modeled by the Euclidean-vertex model where $\dim\calH=3$. To keep the statistical properties of the data approximately invariant, we only make use of the data in the winter months December, January, February and March, which contains a total of 114 days of records. Since the original data has missing values, we first omit those sensors with more than $10\%$ missing values so that $n=30$ sensors are left. Then we fill in the remaining missing values by taking the average of the nearest two days' records. 
	
	We embed the sensors in a weighted graph using their geographical coordinates and a $K$-NN method as in \cite{LouPer:J19}. To be specific, we employ the 5-NN strategy to determine the connectivity between vertices. Next, we accord each edge the weight $\exp(-d(i,j)^2/\sigma^2)$, where $d(i,j)$ is the geographic distance between sensors $i$ and $j$, and $\sigma^2$ the variance of all distances between pairs of sensors. 
	
	We randomly choose 57 days' records as the training set. The remaining days' records form the test set. We randomly remove some data, and then try to recover these values. To this end, we use the training set to estimate the JPSD and covariance operators as described at the beginning of \cref{sect:num_exp}, then apply the Wiener filter to recover the missing values in the test set. 
	In this experiment, we take the normalized error
	\begin{align*}
		\frac{\bbE\norm{\hat{X}-X}}{\bbE\norm{X}}
	\end{align*}
	as the performance metric.
	
	The dataset is indexed by three dimensions: vertex, feature and time in hourly intervals. To illustrate the effectiveness of the \gls{JWSS} assumption, we test different frameworks under two missing data models:
	\begin{enumerate}
		\item Consecutive missing model. For each day in the dataset, we randomly choose a set of (vertex, feature) indices, and remove their corresponding data during a time period whose length is a geometric random variable. This model reflects the original dataset, which contains missing values over a continuous time period. Since both training and test sets have missing values, we test two methods for the TV framework to recover \gls{JPSD}: either zero-pad missing values, or perform linear interpolation to recover the training set first and then estimate \gls{JPSD}. 
		Both TV and GSP process the three features separately.
		\item Uniform missing model. For each day in the test set, we randomly choose a set of (vertex, feature, time) indices and remove their data. In this model, the training set contains no missing values. Therefore, for \gls{GRP}, we use the Cartesian product of the spatial graph and time graph (the cyclic graph with 24 vertices) as the underlying graph. On this graph, each vertex contains three features.
		In this case, applying TV is equivalent to applying GSP on this product graph, in which the three features are processed separately.
	\end{enumerate}
	
	\begin{figure}[!htbp]
		\centering
		\begin{subfigure}[b]{0.9\columnwidth}
			\centering
			\includegraphics[width=\linewidth, trim=.5cm 0cm 1cm .5cm, clip]{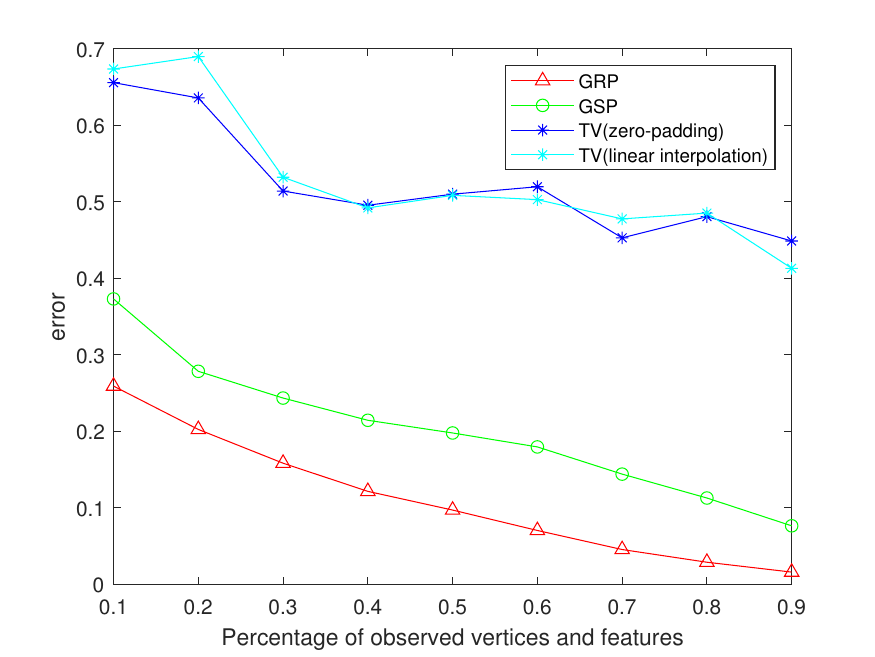}
			\caption{Consecutive missing model}
			\label{fig:contmis}
		\end{subfigure}
		\\
		\begin{subfigure}[b]{0.9\columnwidth}
			\centering
			\includegraphics[width=\linewidth, trim=.5cm 0cm 1cm .5cm, clip]{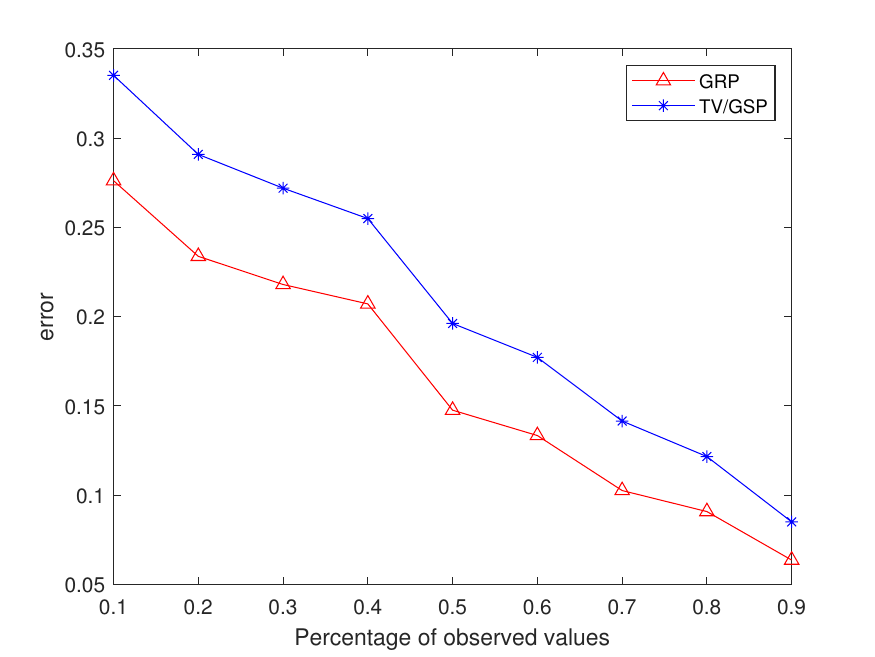}
			\caption{Uniform missing model}
			\label{fig:unimis}
		\end{subfigure}
		\caption{Recovery performance by Wiener filters under different frameworks. In \cref{fig:contmis}, the lengths of missing periods are generated by $\text{Geo}(1/12)$. Each point is the average of 40 repetitions in \cref{fig:contmis}, and 50 repetitions in \cref{fig:unimis}.}
		\label{fig:krakowair:casualwf}
	\end{figure}
	
	From \cref{fig:contmis}, we observe that the \gls{GRP} Wiener filter dominates the traditional GSP Wiener filter and TV Wiener filter under the consecutive missing model. 
	This is due to the fact that the \gls{GRP} incorporates the correlation between features via \cref{eq:exp:H-V_C_X_estimator} while the other methods regard them as independent. 
	In fact, the features we use (i.e., PM1, PM2.5 and PM10) are strongly correlated as indicated by their correlation coefficient matrix estimated empirically from the whole dataset:
	\begin{align*}
		\begin{pmatrix}
			1.0000 &   0.9954  &  0.9898\\
			0.9954  &  1.0000  &  0.9955\\
			0.9898  &  0.9955  &  1.0000\\
		\end{pmatrix}.
	\end{align*}

	Furthermore, the consecutive missing values along the time domain in the training set cause difficulties for TV \gls{JPSD} estimation. Since the \gls{GRP} model has the flexibility to ignore the missing period and to use only the complete data to learn \gls{JPSD}, it is more accurate than TV methods.
	
	From \cref{fig:unimis}, we observe that the \gls{GRP} Wiener filter outperforms the TV Wiener filter, which is also due to the incorporation of correlation between features in \gls{GRP}.
	
	\subsection{Continuous-time signal recovery}\label{subsect:contrec}
	
	In this subsection, we evaluate the recovery performance on continuous-time graph signals under the \gls{GRP} framework, and compare it with both TV and the traditional time stationary~(TS) framework. The graph is generated by the Erd\H{o}s-R\'{e}nyi model with $30$ vertices and edge probability $0.5$. We enforce the graph to be connected. The generalized graph signal is generated as a randomized linear combination of the tensor products of the graph Fourier basis and sinusoids:
	\begin{align}\label{eq:genercont}
		X(t) = \sum_{k,l} d_{kl}\phi_k\otimes \sin(\beta_lt),~t\in[-\pi,\pi],
	\end{align}
	where $d_{kl}$ are independently generated by the Gaussian distribution $\dist{\calN}[0,\sigma^2_{k,l}]$, and $\beta_l$ are real-valued constants. We note that $X(t)$ is not bandlimited as long as there exists a $\beta_l$ that is not integer. Both training and test sets consist of noisy observations sampled from different vertices at different time instances. 
	
	To recover the continuous-time signal, we first apply the variational EM algorithm on the training set to estimate the \gls{PSD} and noise power. To be specific, no matter what framework we use, the observation model can be written as
	\begin{align}\label{eq:regmodel}
		y = \bB c+e,
	\end{align}
	where each column of $\bB$ contains the values of a basis function at the sample points, and $c=(c_1,c_2,\ldots,c_s)\T$ are the Fourier coefficients that are assumed to be independent and have distribution $\dist{\calN}[0,p_i]$. The observation error vector $e$ has components independently generated via $\dist{\calN}[0,\sigma^2]$.  
	
	For instance, in the \gls{GRP} framework, the basis $\set{f_{k,\tau}}$ is $\set{\phi_k\otimes\sin(\tau t) \given k=1\ldots,n,\ \tau\in\bbN}\cup\set{\phi_k\otimes\cos(\tau t) \given k=1\ldots,n,\ \tau\in\bbN}$. In practice, we only use a subset of them so that  $1\leq \tau\leq m_0$ for some $m_0$. Suppose the sampled time instances on the $i$th vertex are $\set{t_{i1},t_{i2},\ldots,t_{iq}}$. Then for each basis function $f\in\set{f_{k,\tau}}$, its values at the sample points can be written as a vector
	\begin{align}\label{eq:basefunc}
		b=(f_1(t_{11}),\ldots,f_1(t_{1q}),\ldots,f_n(t_{n1}),\ldots,f_n(t_{nq}))\T,
	\end{align}
	where $f_i~(i=1,\cdots,n)$ denotes the $i$-th vertex signal of $f$. (Recall that $f(t)$ is an $n$-dimensional vector valued function). By concatenating all $b$ into a matrix we obtain the matrix $\bB$. 
	
	For the TV framework, we use the basis $\set{\phi_k\otimes\psi_\tau \given k=1,\ldots,n,\ \tau=1,\ldots,m_0}$, where $\set{\psi_{\tau}}$ denotes the Fourier basis induced by a cyclic graph. After recovering all values on the time grid, we apply linear interpolation to recover the continuous signal. Under the time stationary assumption, every vertex signal is stationary in the time domain. Each vertex signal is processed separately, with model \cref{eq:regmodel} and basis functions $\set{\sin(\tau t)\given \tau=1,\ldots,m_0}\cup\set{\cos(\tau t)\given \tau=0,\ldots,m_0}$.
	
	Now that $y$ and $\bB$ are known, we apply the EM algorithm to solve for the estimates of $\set{p_i}$ and $\sigma^2$. However, since the \gls{PSD} values $\set{p_i}$ are not assumed to be equal, the posterior $p(c,\sigma^2,p_1,\ldots,p_s\mid y)$ cannot be computed analytically. Therefore, we apply variational EM \cite{TziLikGal:J08}, which admits explicit analytical optimal values at each iterative step, to estimate these hidden values.
	
	On the test set, the continuous-time signals are recovered based on the sample values and the information learned from the training set. Since we have assumed the regression model \cref{eq:regmodel} in which $\set{c_i}$ and $e$ are normal random variables, it suffices to use the posterior mean as the estimator of the Fourier coefficients.
	
	In this experiment we consider two sampling schemes:
	\subsubsection{Equally spaced sampling}\label{subsubsect:equitime}
	The samples are collected from a subset of equally spaced points of $[-\pi,\pi]$. To be specific, on each vertex we randomly choose $m$ points from the time grid $\set{t_i \given t_i=-\pi+2(i-1)\pi/(2m-1),\ i=1,\ldots,2m}$. The graph signals are generated via \cref{eq:genercont} given $\tilde{\beta}=(\beta_1,\beta_2,\beta_3)$, and
	$\sigma^2_{k,l}=10/(kl)$ such that the signals are smooth over the graph. 
	
	We investigate the performance of different frameworks under varying grid density (represented by the number of samples $m$) and noise energy (represented by SNR in dB). For each fixed pair of ($m$, SNR), we uniformly generate $100$ $\tilde{\beta}$ vectors via $\dist{Unif}[1,15]$. Then, for each $\tilde{\beta}$, we generate a training set and test set, each containing 60 realizations of continuous data from \cref{eq:genercont}. To measure the recovery performance, for each $\tilde{\beta}$, we compute the relative error
	\begin{align}\label{eq:relerr}
		\frac{\bbE\norm{\hat{X}-X}}{\bbE\norm{X}},
	\end{align}
	where the expectation denotes averaging over the test set.
	
\begin{figure}[!htbp]
		\centering
		\begin{subfigure}[b]{0.9\columnwidth}
			\centering
			\includegraphics[width=\linewidth, trim=1cm 0cm 1cm .5cm, clip]{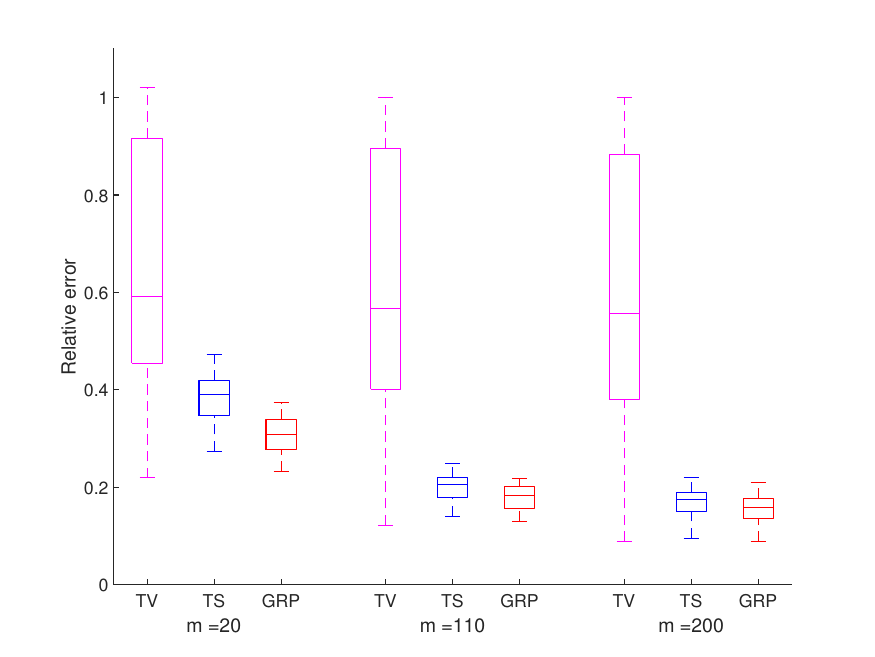}
			\caption{Performance under varying time grid density.}
			\label{fig:equisampN}
			
		\end{subfigure}\\
		\begin{subfigure}[b]{0.9\columnwidth}
			\centering
			\includegraphics[width=\linewidth, trim=1cm 0cm 1cm .5cm, clip]{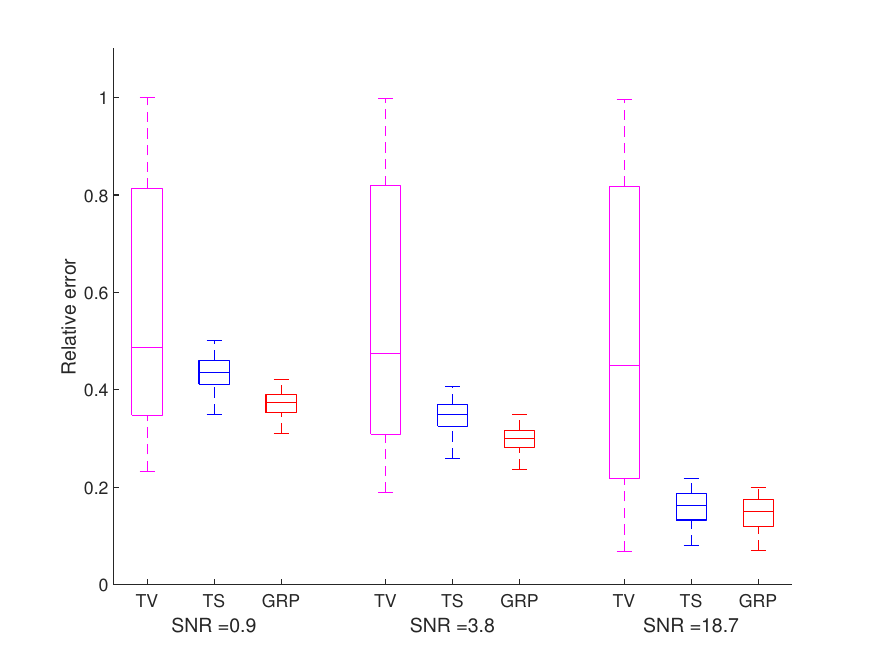}
			\caption{Performance under varying noise energy.}
			\label{fig:equinois}
			
		\end{subfigure}
		\caption{Boxplots of recovery performance under different frameworks via equally spaced sampling. In \cref{fig:equisampN}, the SNR is 8.6. In \cref{fig:equinois}, $m=60$.}
		\label{fig:equisamp}
	\end{figure}

	We summarize the results as box plots (\cref{fig:equisamp} and \cref{fig:unisamp}), where each box reflects the distribution of recovery error with $100$ different $\tilde{\beta}$. To recover the test signal, we set $m_0=20$. 
	From \cref{fig:equisamp}, we observe that \gls{GRP} outperforms both TS and TV. 
	Compared with pure time domain based methods like TS, GRP makes use of the graph structure, which provides additional information.
	In addition, \gls{GRP} uses a basis of continuous functions, while the interpolation step in TV fails to recover the high-frequency variations in the signal. Therefore, the performance of TV is largely affected by the choice of $\tilde{\beta}$.
	
	\subsubsection{Uniformly distributed sampling}
\begin{figure}[!htbp]
		\centering
		\begin{subfigure}[b]{0.9\columnwidth}
			\centering
			\includegraphics[width=\linewidth, trim=1cm 0cm 1cm .5cm, clip]{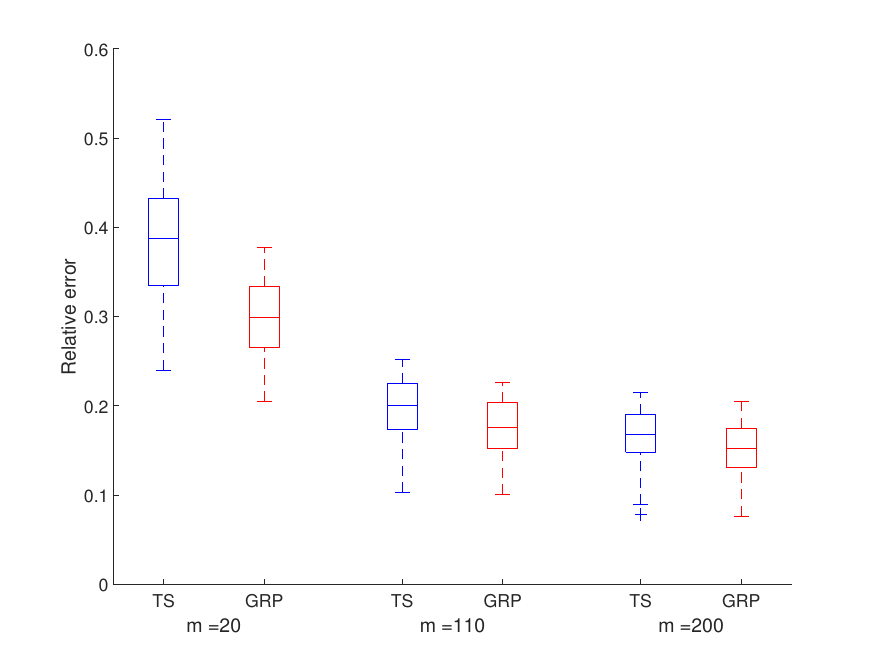}
			\caption{Performances under varying number of samples.}
			\label{fig:unisampN}
		\end{subfigure}\\
		\begin{subfigure}[b]{0.9\columnwidth}
			\centering
			\includegraphics[width=\linewidth, trim=1cm 0cm 1cm .5cm, clip]{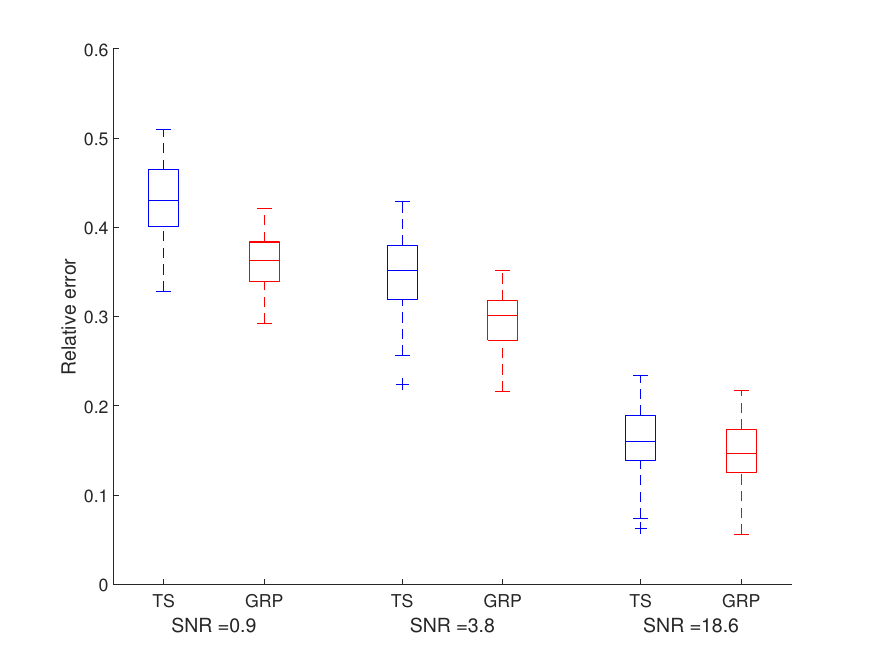}
			\caption{Performances under varying noise energy.}
			\label{fig:uninois}
		\end{subfigure}
		\caption{Boxplots of recovery performance under different frameworks via uniformly distributed sampling on $[-\pi,\pi]$. In \cref{fig:unisampN}, the SNR is 8.6. In \cref{fig:uninois}, the number of samples $m$ on each vertex is 60.}
		\label{fig:unisamp}
	\end{figure}

	In this setting, the sampled time instances on each vertex are uniformly distributed on $[-\pi,\pi]$. Since the time instances are not placed in a time grid, the TV framework cannot be applied in this setting. The graph signals are generated the same way as \cref{subsubsect:equitime}. The performance is measured by \cref{eq:relerr}. We investigate the performance of different frameworks under varying number of samples and noise energy. We observe from \cref{fig:unisamp} that \gls{GRP} outperforms TS.
	
	\subsection{Comparing sampling strategies}\label{subsect:MSEcomp}
	In this subsection, we illustrate the usefulness of \cref{thm:mse} in comparing the \gls{MSE}s of different sampling methods. To this end, we first estimate the \gls{JPSD} from the training set. On the test set, we randomly generate sampling sets with different sizes. We compute the theoretical \gls{MSE} values for all the sampling sets via \cref{eq:theomse}, and compare them with the empirical \gls{MSE}. If the empirical \gls{MSE} is close to the theoretical value (computed based on the estimated JPSD), this experiment demonstrates that it is possible to estimate the quality of a sampling set. 
	
\begin{figure}[!htbp]
		\centering
		\includegraphics[width=\linewidth]{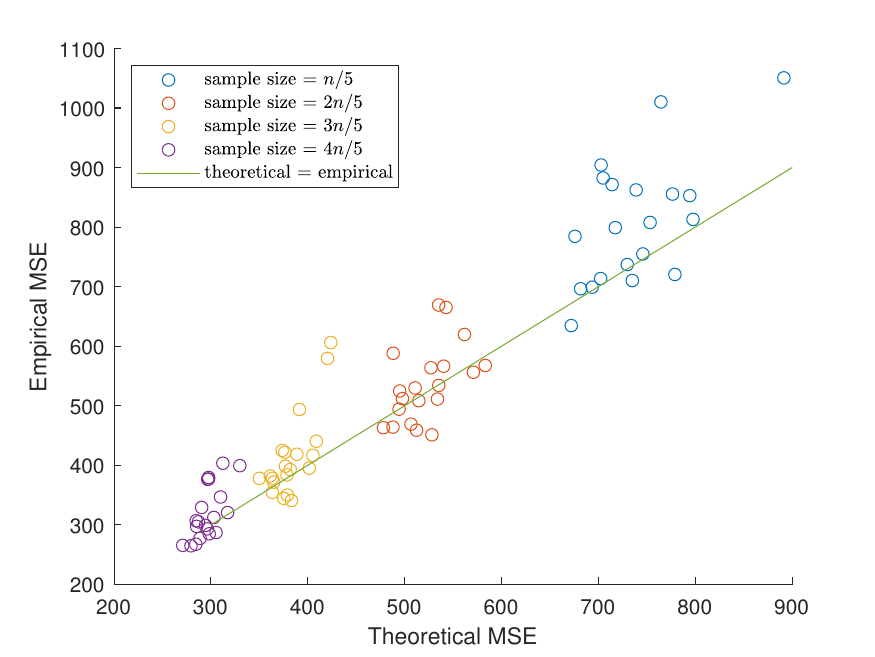}
		\caption{Theoretical and empirical recovery performances under different sample sets. Each point represents the theoretical and empirical errors of a sample set.}
		\label{fig:mse}
	\end{figure}

	The experiment is done on the Molene weather dataset\footnote{\url{https://donneespubliques.meteofrance.fr/donnees_libres/Hackathon/RADOMEH.tar.gz}} published by the French national meteorological service. It contains hourly weather records in the region of Brest, France in January 2014. The temperature records we use are measured by 32 stations. We split the data into 31 periods, so that each period contains 24 hourly records. The graph is constructed similarly as in \cref{subsec:exp:causal_WF}: we use $5$-NN to connect the vertices according to their geographic distances $d(i,j)$, and assign the edges with weight $\exp(-d(i,j)^2/\sigma^2)$. Here, $\sigma^2$ denotes the variance of all distances between pairs of sensors. We randomly sample 20 days' records among 31 days as the training set, and the remaining 11 days' records form the test set. We aim to recover the test set from noisy observations on the sampled vertices. The noise energy is set to be $1/5$ of the signal energy. In order to simulate different sampling strategies, we randomly generate the sample sets $\scU\subset V$ with different sizes $(1/5,2/5,3/5,4/5)n$. For each sample size, we generate 20 sample sets. To recover the original signal and compute the theoretical \gls{MSE}, we employ the Euclidean-vertex model. We observe from \cref{fig:mse} that the empirical \gls{MSE} is aligned with the theoretical \gls{MSE}, hence \cref{eq:theomse} can be utilized to measure the quality of sampling sets.
	
	\section{Conclusion}\label{sect:conclusion}
	
	We have introduced the concepts of a \gls{GRP} and \gls{JWSS} processes. These concepts generalize the existing stationarity models for GSP to the more general GGSP setting. Concrete cases such as the time-vertex model, continuous-time graph signal, and multichannel graph signal are encompassed in this framework. The stationarity models in the graph and Hilbert space domains are related in the sense that JWSS implies wide-sense stationarity in the respective domains. The explicit and approximate forms of Wiener filters for denoising and signal completion are also derived. The implementation of the \gls{GRP} framework is illustrated via several numerical experiments. In the case of a finite-dimensional feature space, the shift operator in the Hilbert space can be learned from data without prior knowledge. When dealing with finite samples of \gls{GRP}s taking values in a Hilbert space that may be infinite-dimensional, a variational EM algorithm is proposed to simultaneously estimate the PSD and noise energy, allowing the recovery of continuous signals with finite and noisy observations.
	
	An alternative approach to inference for graph signals is machine learning data-driven methods like \gls{GNN}s. For example, \gls{GAE} \cite{DoMinDel:C20} has been used to denoise graph signals. In general, the training complexity of \gls{GRP} is lower than \gls{GNN} and a \gls{GRP} model has fewer parameters than \gls{GNN}. For \gls{GRP}, it suffices to determine the filter coefficients and bandwidth, while a \gls{GNN}'s weights depend on the number of layers, and the number of channels in different layers. In addition, the Wiener filter for \gls{GRP} has an explicit solution \cref{eq:wienerco}, while a \gls{GNN} depends on training data to optimize its weights. The Wiener filters we have proposed are linear transformations, while for a \gls{GNN}, the test complexity depends on the number of hidden layers, feature dimension and connection settings. Finally, it is easier to interpret \gls{GRP} as it is a statistical model. On the other hand, GNNs can learn nonlinear relationships and do not require the assumption of JWSS. The pros and cons of statistical parametric models versus data-driven models are expounded in \cite{Bre:J01,BzdAltKrz:J18}. Comparison of GRP and GNN in different applications is an interesting future research direction.
	\appendices
	
	\section{Proof of \cref{thm:contstime} and \cref{thm:traceeq}}
	\label[Appendix]{sect:proof:thm1and2}
	
	This appendix contains the proofs of \cref{thm:contstime,thm:traceeq}, which assume $\calH=L^2(\Upsilon)$. The proofs generalize the ones given in \cite{HsiEub:15}, which assumes a one-dimensional stochastic process. The condition of continuity of trajectories and compactness of $\Upsilon$ are also relaxed here compared to \cite{HsiEub:15}.
	
	\begin{IEEEproof}[Proof of \cref{thm:contstime}]
		From \cref{prop:Xmeasurable} in the supplementary and \cref{thm:contstime:L2}, it suffices to show that for any $y\in\CH$, the map $\angles{X(\omega,\cdot),y}:\Omega\mapsto\Complex$ is measurable. Because $\bbC^n$ is finite dimensional, it suffices to consider $y$ such that for each $t\in\Upsilon$, $y(t)=y(\cdot,t)\in\bbC^n$.
		From Cauchy–Schwarz inequality, we have
		\begin{align}
			&\int_{\Upsilon}\int_{\Omega} |y(t)^*X(\omega,t)| \ud\mu(\omega)\ud\nu(t)\nn
			&\leq \int_{\Upsilon}\int_{\Omega} \norm{X(\omega,t)}\norm{y(t)} \ud\mu(\omega)\ud\nu(t)\nn
			&=\int_{\Upsilon}\bbE[\norm{X(t)}]\norm{y(t)}\ud\nu(t)\nn
			&\leq \parens*{\int_{\Upsilon}\bbE[\norm{X(t)}]^2\ud\nu(t)}^{1/2} \parens*{\int_{\Upsilon}\norm{y(t)}^2\ud\nu(t)}^{1/2}\nn
			&\leq \parens*{\int_{\Upsilon}\bbE[\norm{X(t)}^2]\ud\nu(t)}^{1/2} \parens*{\int_{\Upsilon}\norm{y(t)}^2\ud\nu(t)}^{1/2}. \label{ipXybound}
		\end{align}
		From \cref{thm:contstime:mc}, we have
		\begin{align*}
			\int_{\Upsilon}\bbE[\norm{X(t)}^2]\ud\nu(t)=\int_{ \Upsilon}\tr(\bK_X(t,t))\ud\nu(t)<\infty.
		\end{align*}
		In addition, since $y(v,\cdot) \in L^2(\Upsilon)$ for each $v\in \scV$, the \gls{RHS} of \cref{ipXybound} is finite. By Fubini's theorem, we obtain that the following integral, as a function of $\omega$, 
		\begin{align*}
			\angles{X(\omega,\cdot), y} = \int_{\Upsilon} y(t)^*X(\omega,t) \ud\nu(t)
		\end{align*}
		is measurable (in fact, integrable). This concludes the proof.
	\end{IEEEproof}
	
	\begin{IEEEproof}[Proof of \cref{thm:traceeq}]
		We first verify that the pointwise mean defined in \cref{thm:contstime}, $m_X(\cdot)\in \Complex^n\otimes L^2(\Upsilon)$ as follows:
		\begin{align*}
			\int_{\Upsilon} \norm{m_X(t)}^2\ud\nu(t)
			&\leq \int_{\Upsilon}\bbE[\norm{X(t)}^2]\ud\nu(t)\\
			&=\int_{\Upsilon}\tr(\bK_X(t,t))\ud\nu(t)\\
			&<\infty.
		\end{align*}
		We then have for any $u\in\CH$ with $u(t)\in\bbC^n$ written in the form \cref{vector-form},
		\begin{align*}
			\angles{m_X(\cdot),u(\cdot)}&=\int_{\Upsilon} u(t)^*m_X(t)\ud\nu(t)\\
			&=\int_{\Upsilon} \int_{\Omega} u(t)^*X(\omega,t) \ud\mu(\omega)\ud\nu(t).
		\end{align*}
		Using the same argument as in the proof of \cref{thm:contstime},  the integrals can be interchanged, so that 
		\begin{align*}
			\angles{m_X(\cdot),u(\cdot)}
			&=\int_{\Omega}\int_{\Upsilon} u(t)^*X(\omega,t) \ud\nu(t)\ud\mu(\omega)\\
			&=\bbE[\angles{X,u}],
		\end{align*}
		which is the definition of the mean element of $X$ in \cref{def:mX}.
		
		In the rest of the proof, without loss of generality, we assume $m_X=0$. By definition \cref{def:CX}, we have for any $f, g\in\CH$,
		\begin{align}
			\angles{\bC_X f,g}
			&=\bbE[\overline{\angles{X,f}}\angles{X,g}]\\
			&=\bbE[\int_{\Upsilon\times \Upsilon}f(t)\T\bar{X}(\cdot,t)X(\cdot,s)\T\bar{g}(s)\ud\nu(t)\ud\nu(s)].\label{CXfg}
		\end{align}
		In the proof of \cref{thm:contstime}, we have shown that the function $\norm{X(\omega,t)}\norm{y(t)}$ is integrable on $\Upsilon$. We use this fact with Fubini's theorem to interchange the expectation and integral in \cref{CXfg} to obtain
		\begin{align*}
			\angles{\bC_X f,g}
			&=\int_{\Upsilon\times \Upsilon} f(t)\T\widebar{\bK}_X(t,s)\bar{g}(s)\ud\nu(t)\ud\nu(s)\\
			&=\angles*{\int_{\Upsilon}\bK_X(s,t)f(t)\ud\nu(t),g(s)}.
		\end{align*}
		Note that the first element in the above inner product is an integral operator on $f(t)$, therefore $\bC_X$ coincides with this operator by definition.
		
		To prove the second part of \cref{thm:traceeq}, we employ the generalized Mercer's theorem in terms of a matrix-valued kernel \cite{VitUmaVil:J13}. From \cite[Theorem A.1]{VitUmaVil:J13}, there exists a sequence $\set{f_i(t)\given i=1,2,\ldots}\subset\Complex^n\otimes L^2(\Upsilon)$ such that 
		\begin{enumerate}
			\item $\bC_X f_i = \sigma_i f_i$ with $\sigma_i>0$;
			\item $\{f_i(t)\}$ forms an orthonormal basis of $\ker(\bC_X)^\perp=\overline{\ima(\bC_X)}$;
			\item $\{f_i(t)\}\subset \calH_K$, where $\calH_K$ is the reproducing kernel Hilbert space induced by the kernel $\bK_X(s,t)$.
		\end{enumerate}
		Using \cite[Remark 3.3]{VitUmaVil:J13}, since $\{f_i(t)\}\subset \calH_K$, $f_i(t)$ are also continuous \gls{wrt} the topology induced by $\bK_X(s,t)$. Then according to \cite[Theorem 3.4]{VitUmaVil:J13}, the kernel function $\bK_X(s,t)$ can be decomposed as follows
		\begin{align*}
			\bK_X(s,t)=\sum_{i=1}^{\infty} \sigma_i \bar{f}_i(s)f_i(t)\T,
		\end{align*}
		for all $s,t\in \Upsilon$ except on a zero-measure set. Let $f_i(t)=(f_i^{(1)}(t),\ldots,f_i^{(n)}(t))\T$. Therefore, we can compute the integral of $\tr(\bK_X(t,t))$ as
		\begin{align*}
			\int_{\Upsilon}\tr(\bK_X(t,t))\ud\nu(t)&=\int_{\Upsilon}\sum_{l=1}^{n}\sum_{i=1}^{\infty}\sigma_i|f_i^{(l)}(t)|^2\ud\nu(t)\\
			&=\int_{\Upsilon}\sum_{i=1}^{\infty}\sum_{l=1}^{n}\sigma_i|f_i^{(l)}(t)|^2\ud\nu(t)\\
			&=\sum_{i=1}^{\infty}\int_{\Upsilon}\sigma_i\norm{f_i(t)}^2\ud\nu(t)\\
			&=\sum_{i=1}^{\infty}\sigma_i\\
			&=\tr(\bC_X),
		\end{align*}
		which concludes the proof.
	\end{IEEEproof}

\section{A Brief Introduction To Hilbert Spaces And The Bochner Integral}
\label{sect:Hspace_intro}
In this section, we provide a brief overview of some concepts related to Hilbert spaces and the Bochner integral. Readers are referred to \cite{HsiEub:15,Bak:J73,Vak:87,DebMik:00} for further details.

A Hilbert space $\calH$ is a complete normed space (i.e., Banach space), whose norm $\norm{\cdot}$ is induced by an inner product $\angles{}$ so that $\norm{x}=\sqrt{\angles{x,x}}$. Examples of Hilbert spaces include $L^2(\Upsilon)$ and the Euclidean spaces. Here $\Upsilon$ can be any measure space. 

The norm $\norm{\cdot}$ naturally induces a topology on $\calH$ whose topological basis consists of the open balls centered at each $y\in\calH$ with radius $\delta>0$, denoted as
\begin{align*}
	B(y,\delta)=\set{x\in\calH \given \norm{x-y}<\delta}.
\end{align*}
This topology defines the Borel $\sigma$-algebra $\calB$ of $\calH$ as the smallest $\sigma$-algebra containing all the open subsets of $\mathcal{H}$, so that $(\calH,\calB)$ becomes a measurable space. 

Two Hilbert spaces $\calH_1$ and $\calH_2$ are said to be isomorphic ($\calH_1\cong\calH_2$) if there exists a bijective map $\psi:\calH_1\mapsto\calH_2$ such that:

\begin{align*}
	\psi(ax+by)&=a\psi(x)+b\psi(y),\\
	\angles{\psi(x),\psi(y)}&=\angles{x,y},
\end{align*}
for arbitrary $a,b\in\Complex$, $x,y\in\calH_1$. 

An operator $\bC$ on $\calH$ is a linear map from $\calH$ to itself. In this paper we only consider bounded operators, i.e.,
\begin{align*}
	\sup_{\norm{x}=1} \norm{\bC (x)} <\infty.
\end{align*}
An operator $\bC$ is bounded if and only if it is continuous. When $\dim\calH=d<\infty$, an operator can be represented as a $d\times d$ matrix $\bC$. Symmetric matrices form a notable class of matrices in linear algebra, since its eigenvectors form an orthonormal basis for the whole space. The counterparts of these in Hilbert spaces are the compact self-adjoint operators, defined as follows.

\begin{Definition}\label{def:compact_operator}
	An operator $\bC$ is \emph{compact} if for any bounded sequence $(x_k)_{k\geq1} \subset \calH$, there exists a subsequence $(x_{k_i})_{i\geq1}$ such that $(\bC (x_{k_i}))_{i\geq1}$ converges. 
\end{Definition}

\begin{Definition}\label{def:selfadjoint_operator}
	The operator $\bC^*$ defined by
	\begin{align*}
		\angles{u,\bC^* (v)}=\angles{\bC (u),v}
	\end{align*}
	for all $u,v\in\calH$ is called the \emph{adjoint operator} of $\bC$. An operator $\bC$ is called \emph{self-adjoint} if $\bC=\bC^*$.
\end{Definition}

If an operator $\bC$ satisfies both \cref{def:compact_operator} and \cref{def:selfadjoint_operator}, then it admits an eigendecomposition.

\begin{Proposition}\label{prop:spectral_decomp}
	\cite[Corollary 4.10.2, 4.10.3]{DebMik:00} 
	Let $\bC$ be a compact self-adjoint operator on $\calH$. Then $\calH$ has an orthonormal basis $\{\psi_k\}_{k=1}^\infty$ consisting of the eigenvectors of $\bC$. Furthermore, we have
	\begin{align*}
		\bC (x)=\sum_{k=1}^\infty \lambda_k\angles{x,\psi_k}\psi_k.
	\end{align*}
	In other words, $\bC$ can be decomposed into finite-rank projection operators:
	\begin{align*}
		\bC=\sum_{k}^\infty \lambda_k\bPi_{\psi_k}.
	\end{align*}
\end{Proposition}

\begin{Definition}\label{def:trace_class}
	An operator $\bC$ on $\calH$ is \emph{trace-class} if  
	\begin{align*}
		\tr(\bC)=\sum_{k=1}^\infty \angles*{(\bC^*\bC)^{1/2} e_k,e_k}
	\end{align*}
	converges for some orthonormal basis $\{e_k\}_{k=1}^\infty$ of $\calH$. $\tr(\bC)$ is known as the \emph{operator trace} of $\bC$.
\end{Definition}
It can be shown that \cref{def:trace_class} is independent of the choice of orthonormal basis $\{e_k\}_{k=1}^\infty$.

For a function taking values in a Hilbert space, its Bochner integral is defined by the limit of a series of simple functions, which is similar to the definition of the Lebesgue integral. Let $(\Omega,\calF,\mu)$ be a measure space. In the following, we define the Bochner integral of a measurable function $f:\Omega \mapsto \calH$. We start with the definition of a simple function.

\begin{Definition}\label{def:simpfunc}
	A \emph{simple function} $g$ is a linear combination of characteristic functions:
	\begin{align*}
		g=\sum_{k=1}^n x_k\indicator{A_k},
	\end{align*}
	wherein $A_k\in\calF$, $A_k\cap A_l=\varnothing$ for $k\neq l$, and $x_k\in\calH$, for all $k\geq1$. The Bochner integral of $g$ is defined to be 
	\begin{align*}
		\int_{\Omega}g\ud \mu = \sum_{k=1}^n\mu(A_k)x_k \in\calH.
	\end{align*}
\end{Definition}

\begin{Definition}\label{def:Bochnerint}
	A measurable function $f:\Omega\mapsto\calH$ is Bochner integrable if there exists simple functions $\{f_k\}_{k=1}^\infty$ such that
	\begin{align*}
		\lim_{k\rightarrow\infty}\int_{\Omega} \norm{f-f_k}\ud \mu=0.
	\end{align*}
	Its integral on any set $E\in\calF$ is defined as 
	\begin{align*}
		\int_{E} f \ud \mu:=\lim_{k\rightarrow\infty}\int_{E} f_k \ud \mu.
	\end{align*}
\end{Definition}
It can be shown that \cref{def:Bochnerint} is independent of the sequence of converging simple functions chosen. One may also note that neither \cref{def:simpfunc} nor \cref{def:Bochnerint} directly makes use of the inner product but only the norm. In fact, the Bochner integral is also applicable to functions taking values in Banach spaces. Alternative definitions and properties of Bochner integral can be found in \cite{Yos:80,Mik:78,Mik:14}.

\section{Random Elements in a Hilbert Space}\label{sect:randHilbert}

In this section, we introduce the definition of a random element and its moments. We provide interpretation of a random element's moments via the Bochner integral, and explain why they are the generalizations of the mean and covariance in a Euclidean space to a Hilbert space.

Consider a probability space $(\Omega,\calF,\mu)$, where $\Omega$ is a sample space, $\calF$ a $\sigma$-algebra and $\mu$ a probability measure. Let $\calH$ be a complex separable Hilbert space with inner product $\ip{}{}$. Since $\mathcal{H}$ is a Banach space, it is naturally endowed with the norm-induced topology. This topology defines the Borel $\sigma$-algebra $\calB$ as the smallest $\sigma$-algebra containing all the open subsets of $\mathcal{H}$, so that $(\calH,\calB)$ is a measurable space. A random element is defined as a measurable map $X:\Omega \mapsto \calH$, which induces a probability measure $\bbP$ on $(\calH,\calB)$ given by
\begin{align*}
	\bbP(B)=\mu(X^{-1}(B)),~\forall B\in \calB.
\end{align*}
A sufficient and necessary condition for $X$ to be measurable (i.e., $X^{-1}(B) \in \calF$ for all $B\in\calB$) is as follows.

\begin{Proposition}\label{prop:Xmeasurable}
	\cite[Theorem 7.1.2]{HsiEub:15} $X$ is measurable if and only if $\angles{X,u}$ is measurable for all $u\in\calH$.
\end{Proposition}
\cref{prop:Xmeasurable} enables us to verify the measurability of $X$ by investigating that of a family of complex-valued functions. In \cref{sect:statmodel}, we utilize \cref{prop:Xmeasurable} to model continuous-time graph processes as random elements.

In the paper, \cref{assumpt:mean_norm} ensures the existence of a random element's mean and covariance operators: 
\begin{align}
	\angles{m_{X},u}
	&=\bbE[\angles{X,u}] =\int_{\calH} \angles{\xi,u} \ud\bbP(\xi), \label{sdef:mX}\\
	\angles{\bC_{X}u, v}
	&=\bbE[\overline{\angles{X-m_X,u}}\angles{X-m_X,v}]\nn
	&=\int_{\calH} \overline{\angles{\xi-m_X,u}}\angles{\xi-m_X,v} \ud\bbP(\xi). \label{sdef:CX}\\
\end{align}

Suppose two random elements $X:\Omega\mapsto\calH_1$ and $Y:\Omega\mapsto\calH_2$ induce a joint probability measure $\bbP_{XY}$ on $\calH_1\times\calH_2$, and $\bbE\norm{(X,Y)}^2<\infty$, their cross-covariance operator is given by
\begin{align}
	\angles{\bC_{XY}u,v}&=\bbE[\overline{\angles{Y-m_Y,u}}\angles{X-m_X,v}]\nn
	&=\int_{\calH_1\times\calH_2}\overline{\angles {\zeta-m_Y,u}}\angles{\xi-m_X,v} \ud\bbP_{XY}(\xi,\zeta). \label{sdef:CXY}
\end{align}

Let $x\otimes_1y$ denote the operator $\bA$ that maps $u$ to $\bA u = \angles{u,y}x$. This is the Kronecker product matrix $\bA=xy^*$ of $x$ and $y$ when both are finite dimensional column vectors. An alternative and more direct way to define the mean element and covariance operator is via the Bochner integral (cf.\ \cref{def:Bochnerint} and \cite{HsiEub:15}):
\begin{align}
	m_X&=\int_{\Omega} X \ud \mu,\\
	\bC_X&=\int_{\Omega} (X-m_X)\otimes_1(X-m_X) \ud \mu.
\end{align}
We can interpret $m_X$ as the (generalized) mean of the random element or vector $X$. For $\bC_X$, we note that $\bbE[(X-m_X)\otimes_1(X-m_X)(u)]=\bbE[\angles{u,X-m_X}(X-m_X)]$, which is nothing but the application of the matrix $\bbE[(X-m_X)(X-m_X)^*]$ to $u$ when $X$ is a finite dimensional random vector in a Euclidean space. It can be shown that covariance operators are bounded, trace-class and self-adjoint (cf.\ \cref{def:trace_class,def:selfadjoint_operator}). 

The cross-covariance operator can also be defined via Bochner integral. If $\bbE\norm{(X,Y)}^2<\infty$, the cross-covariance operator $\bC_{XY}:\calH_2\mapsto\calH_1$ is defined as 
\begin{align}
	\bC_{XY}=\int_{\Omega} (X-m_X)\otimes_1(Y-m_Y) \ud\mu.
\end{align}
Similarly, it can be verified that this definition also degenerates to the standard definitions of mean and covariance of random vectors in finite dimensional Hilbert spaces. By \cref{assumpt:mean_norm}, the covariance and cross-covariance operators are well-defined as Bochner integrals.

Due to the fact that $\bC_X$ is self-adjoint and trace-class, its trace can be computed given an arbitrary orthonormal basis $\{e_k\}_{k=1}^{\infty}$of $\calH$:
\begin{align}\label{eq:operator-trace}
	\tr(\bC_X)=\sum_{k=1}^{\infty} \angles{\bC_X(e_k),e_k}.
\end{align}
The trace of the covariance matrix of a random vector in a finite dimensional space equals to the expectation of its squared norm. Analogously, this fact holds true for a random element, as shown below. \cref{prop:tr=E} is needed in the \gls{MSE} analysis in the main paper.

\begin{Proposition}\label{prop:tr=E}
	For a random element $X$ with $m_X=0$, $\bbE\norm{X}^2=\tr(\bC_X)$.
\end{Proposition}
\begin{IEEEproof}
	Suppose that $\set{e_k}_{k=1}^\infty$ is an orthonormal basis of $\calH$. Then,
	\begin{align*}
		\bbE\norm{X}^2 
		&= \int_{\calH} \sum_{k=1}^{\infty} |\angles{x,e_k}|^2 \ud\bbP(x)\\
		&= \sum_{k=1}^{\infty} \int_{\calH} |\angles{x,e_k}|^2 \ud\bbP(x)\\
		&= \sum_{k=1}^{\infty} \angles{\bC_Xe_k,e_k}\\
		&= \tr(\bC_X),
	\end{align*}
	where the second inequality follows from the monotone convergence theorem, and the third equality from \cref{def:CX}.
\end{IEEEproof}

For a random vector $X$ in a finite dimensional space, one can obtain principal axes by eigendecomposition of its covariance matrix $\bC_X$ to perform PCA. By projecting $X$ onto different principal axes, it is decomposed into uncorrelated components. For a random element $X$ we have the same result as follows.
\begin{Proposition}\label{prop:PCA}
	\cite[Theorem 7.2.6, Theorem 7.2.7]{HsiEub:15} $\bC_X$ admits the eigendecomposition 
	\begin{align*}
		\bC_X=\sum_{i=1}^{\infty} \lambda_i h_i\otimes_1h_i,
	\end{align*}
	where $\set{\lambda_i}_{i=1}^\infty$ are eigenvalues of $\bC_X$ and $\{h_i\}_{i=1}^\infty$ the corresponding orthonormal eigenvectors. The eigenvectors $\{h_i\}_{i=1}^\infty$ form an orthonormal basis for the closure of the image of $\bC_X$, $\overline{\ima \bC_X}$. The random element $X \in \overline{\ima \bC_X}$ almost surely, and can be written as 
	\begin{align*}
		X=\sum_{i=1}^{\infty} \angles{X,h_i}h_i,
	\end{align*}
	where $\{\angles{X,h_i}\}_{i=1}^\infty$ are uncorrelated random variables with zero means and variances $\lambda_i$.
\end{Proposition}
The covariance operator behaves similarly as the covariance matrix under linear transformation. We list a few of their properties here. 

\begin{Proposition}\label{prop:covtrans}
	Suppose $X:\Omega\mapsto\calH_1$ and $Y:\Omega\mapsto\calH_2$ have zero means, and $\bT_1$ and $\bT_2$ are bounded linear operators on $\calH_1$ and $\calH_2$, respectively.
	\begin{enumerate}[(a)]
		\item\label[claim]{ct:X} Let $L_1=\bT_1 X$. Then $\bC_{L_1}=\bT_1\bC_X\bT_1^*$.
		\item\label[claim]{ct:XY} Let $L_2=\bT_2 Y$. Then $\bC_{L_1L_2}=\bT_1\bC_{XY}\bT_2^*$.
		\item\label[claim]{ct:X+Y}  If $X$ and $Y$ are independent, then $\bC_{X+Y}=\bC_X+\bC_Y, \bC_{XY}=\bzero$. 
	\end{enumerate}
\end{Proposition}
\begin{IEEEproof}
	From the definition of a covariance operator, for any $u,v\in\calH$ we have 
	\begin{align*}
		\angles{\bC_{L_1}u,v}&=\bbE[\overline{\angles{L_1,u}}\angles{L_1,v}]\\
		&=\bbE[\overline{\angles{\bT_1X,u}}\angles{\bT_1X,v}]\\
		&=\bbE[\overline{\angles{X,\bT_1^*u}}\angles{X,\bT_1^*v}]\\
		&=\angles{\bC_X\bT_1^*u,\bT_1^*v}\\
		&=\angles{\bT_1\bC_X\bT_1^*u,v},
	\end{align*}
	yielding the result of \cref{ct:X}. The proof of \cref{ct:XY} is similar and omitted here.
	
	For \cref{ct:X+Y}, due to the independence of $X$ and $Y$, we have
	\begin{align*}
		\angles{\bC_{XY}u,v}&=\bbE[\overline{\angles {Y,u}}\angles{X,v}]\\
		&=\overline{\bbE[{\angles {Y,u}}]}\bbE[\angles{X,v}]\\
		&=0.
	\end{align*}
	This implies that $\bC_{XY}=\bzero$. By plugging this result into the definition of $\bC_{X+Y}$ to eliminate the cross terms, we obtain $\bC_{X+Y}=\bC_{X}+\bC_Y$.
\end{IEEEproof}

\section{Linear conditional expectation in Hilbert space}
\label{sect:LCE}
In this section, we present the concept of a \gls{LCE} in Hilbert spaces, and theorems from \cite{KleSprSul:J20} that characterize it. The contents are simplified to fit this paper. Readers are referred to \cite{KleSprSul:J20} for details.

Let $L^2(\Omega,\calF,\mu;\calH)$ be the space of random elements taking value in $\calH$. Consider two random elements $X$ and $Y$ belonging to $L^2(\Omega,\calF,\mu;\calH)$. Let $\mathsf{L}(\calH;\calH)$ be the set of all bounded linear operators on $\calH$. We first define several specific operator spaces:
\begin{align*}
	\mathsf{A}(\calH;\calH)&:=\{\bL:\calH\mapsto\calH | \bL(h)=b+\bA(h) \text{~for some~}\\ &b\in\calH,\bA\in\mathsf{L}(\calH;\calH)\},\\
	\mathsf{L}_Y(\calH,\calH)&:=\{\bL:\calH\mapsto\calH | \bL\text{~is linear and~}\\ &\bL(Y)\in L^2(\Omega,\calF,\mu;\calH)\},\\
	\mathsf{A}_Y(\calH;\calH)&:=\{\bL:\calH\mapsto\calH \given \bL(h)=b+\bA(h) \text{~for some~}\\ & b\in\calH,\bA\in\mathsf{L}_Y(\calH;\calH)\}.
\end{align*}

\begin{Definition}\label{def:LCE}
	The \emph{\gls{LCE}} $\bbE^{\mathsf{A}}[X\given Y]$ is defined to be 
	\begin{align*}
		\bbE^{\mathsf{A}}[X\given Y]=\bPi_{\overline{\mathsf{A}_\calH(Y)}}(X),
	\end{align*}
	wherein $\overline{\mathsf{A}_\calH(Y)}$ is the closure of the space $\set*[\vert]{U \given U=\bL(Y)\text{~for some~}\bL\in\mathsf{A}(\calH;\calH)}$ in $L^2(\Omega,\calF,\bbP;\calH)$. Recall that $\bPi_{\overline{\mathsf{A}_\calH(Y)}}$ denotes the projection operator onto $\overline{\mathsf{A}_\calH(Y)}$.
\end{Definition}
\begin{Proposition}
	$\bbE^{\mathsf{A}}[X\given Y]$ is of the form $\bG(Y)$, where $\bG\in\mathsf{A}_Y(\calH;\calH)$.
\end{Proposition}
If the range inclusion $\ima(\bC_{YX})\subset\ima(\bC_{Y})$ holds, we call it the compatible case.
\begin{Proposition}[Formula for the \gls{LCE}: compatible case]\label{prop:compatible}
	Under the compatible case, the \gls{LCE} has the explicit formula
	\begin{align}
		\bG(y)=m_X+(\bC_Y^\dag\bC_{YX})^*(y-m_Y). \label{eq:LCE_compat}
	\end{align}
\end{Proposition}
\begin{Proposition}\label{prop:cond_for_compat}
	The condition $\ima(\bC_{YX})\subset\ima(\bC_{Y})$ holds if $\ima(\bC_{Y})$ is closed. This condition is trivially met when $\dim\calH<\infty$.
\end{Proposition}

In the non-compatible case, $\bbE^{\mathsf{A}}[X|Y]$ can be approximated by a sequence of finite-rank operators composed with $Y$.
According to \cref{prop:PCA}, suppose $\bC_{Y}$ admits the eigen-decomposition
\begin{align*}
	\bC_{Y}=\sum_{i=1}^{\infty} \lambda_i h_i\otimes_1h_i.
\end{align*}
For every $m\in\bbN$, let $\calH^{(m)}:=\spn\{h_1,\cdots,h_m\}$, and $Y^{(m)}:=\bPi_{\calH^{(m)}}Y$. We define a sequence of operators $\{\bG^{(m)}\}_{m\geq1}$ as 
\begin{align*}
	\bG^{(m)}(y):=m_X+(\bC_{Y^{(m)}}^\dag\bC_{Y^{(m)}X})^*(y-m_Y).
\end{align*}
It can be shown that the sequence $\{\bG^{(m)}(Y)\}_{m\geq1}$ is a good approximation of $\bbE^{\mathsf{A}}[X\given Y]$, as captured in the following result.

\begin{Proposition}[Formula for the \gls{LCE}: incompatible case]\label{prop:incompatible}
	$\{\bG^{(m)}(Y)\}_{m\geq1}$ converges to $\bbE^{\mathsf{A}}[X\given Y]$ in $L^2$ norm, i.e.,
	\begin{align*}
		\bbE\norm*{\bbE^{\mathsf{A}}[X\given Y]-\bG^{(m)}(Y)}^2\to 0,
	\end{align*}
	as $m\to\infty$.
	Let $\bbP_Y$ denote the probability measure induced by $Y$ on $\calH$. Then,
	\begin{align*}
		\norm*{\bG^{(m)}(y)-\bbE^{\mathsf{A}}[X\given Y=y ]}\convas0
	\end{align*}
	$\bbP_Y$-almost surely.
\end{Proposition}

We can define and measure the estimation error of the \gls{LCE} as the \gls{ALCC}.
\begin{Definition}\label{def:ALCC}
	Let $R^{\mathsf{A}}[X\mid Y] := X-\bbE^{\mathsf{A}}[X\given Y]$. The \gls{ALCC} of $X$ given $Y$ is defined as 
	\begin{align*}
		\cov^{\mathsf{A}}_Y[X]:=\bbE[R^{\mathsf{A}}[X|Y]\otimes_1R^{\mathsf{A}}[X|Y]].
	\end{align*}
	In the compatible case, $\cov^{\mathsf{A}}_Y[X]$ can be computed as
	\begin{align*}
		\cov^{\mathsf{A}}_Y[X]=\bC_{X}-\bC_{XY}\bC_{Y}^\dag\bC_{YX}.
	\end{align*}
\end{Definition}

\section{Proof of \cref{thm:mse}}
\label{sect:proof:thmmse}

In this section, we prove \cref{thm:mse}. We start off with a lemma. Recall that $Y$ is restricted to a subset of vertices $\scU$.

\begin{Lemma}\label{lem:C_Ystructure}
	The null space and image space of $\bC_Y$ are respectively 
	\begin{align*}
		\ker \bC_Y&=\overline{\spn}\set{v_i\otimes \psi_{\tau} \given i\in\scU\setcomp~\text{or}~\tau\geq\tau_0}:=V_0,\\
		\ima \bC_Y&=\spn\set{v_i\otimes \psi_{\tau}\given i\in\scU,~\tau<\tau_0}:=V_1.
	\end{align*}
\end{Lemma}
\begin{IEEEproof}
	Note that $\bC_Y=\bPi_\calA(\bC_X+\bC_\calE)\bPi_\calA$. First notice that $\bC_Y(v_i\otimes \psi_{\tau})=0$ when $i\in\scU\setcomp$ or $\tau\geq\tau_0$. Then the continuity of $\bC_Y$ implies that $V_0\subset \ker \bC_Y$. When $i\in\scU$ and $\tau<\tau_0$, we have 
	\begin{align*}
		\bC_Y(v_i\otimes\psi_{\tau})&=\bPi_\calA(\bC_X+\bC_\calE)(v_i\otimes\psi_{\tau})\\
		&=\bPi_\calA(\sum_{l=1}^{\infty}\sum_{k=1}^{n}(p_X(k,l)+p_\calE(k,l))\bPi_{\phi_k\otimes\psi_{l}}(v_i\otimes \psi_{\tau}))\\
		&=\bPi_\calA(\sum_{k=1}^{n}(p_X(k,\tau)+p_\calE(k,\tau))\bm{\Phi}_{ik}\phi_k\otimes\psi_{\tau})\\
		&=\sum_{k=1}^{n}(p_X(k,\tau)+p_\calE(k,\tau))\bPhi_{ik}\bPi_\calA(\phi_k\otimes\psi_{\tau})\\
		&=\sum_{k=1}^{n}(p_X(k,\tau)+p_\calE(k,\tau))\bPhi_{ik}\sum_{j\in\scU}\bPhi_{jk}v_j\otimes \psi_{\tau}\\
		&=\sum_{j\in\scU}(\sum_{k=1}^{n}(p_X(k,\tau)+p_\calE(k,\tau))\bPhi_{jk}\bPhi_{ik})v_j\otimes \psi_{\tau}\\
		&=\sum_{j\in\scU}\bO_{ij}^{(\tau)}v_j\otimes \psi_{\tau},
	\end{align*}
	where $\bO^{(\tau)}=\bPhi_{\scU}\bLambda_\tau\bPhi_{\scU}\T$. In order to simplify notations, we let the row and column indices of $\bO^{(\tau)}$ be consistent with $\scU$. This result indicates that $\ima \bC_Y\subset V_1$. We note that when $\tau<\tau_0$, $\bO^{(\tau)}$ is invertible as a principal submatrix of a positive definite matrix $\bPhi\bLambda_\tau\bPhi\T$. This implies that for $\tau<\tau_0$, when restricted on
	\begin{align}
		\widetilde{V_{\tau}} := \spn\set*{v_i\otimes \psi_{\tau} \given i\in\scU},
	\end{align}
	$\bC_Y$ can be represented as an invertible matrix $\bO^{(\tau)}$. Therefore, the basis of $V_1$ is a subset of $\ima \bC_Y$. Since $\ima \bC_Y$ is closed, it follows that $V_1\subset\ima \bC_Y$, i.e., $\ima \bC_Y=V_1$. Since $V_0=V_1^\perp$ and $\ker \bC_Y=\ima \bC_Y^\perp$, we obtain $\ker \bC_Y=V_0$. The lemma is now proved.
\end{IEEEproof}

We now return to the proof of \cref{thm:mse}.
Define $\hat{X}=\bG(Y)$ as the Wiener filter (i.e., \gls{BLUE}) and  $R:=\hat{X}-X$ as the estimation error. According to \cref{prop:tr=E}, $\bbE\norm{R}^2=\tr(\bC_R)$. In the sequel, we are going to compute it as a function of $\scU$.

From \cref{def:ALCC}, $\bC_R$ equals the \gls{ALCC} $\cov_Y^{\mathsf{A}}[X]$. Therefore, it can be written as  
\begin{align*}
	\bC_R&=\bC_X-\bC_{XY}\bC_Y^\dag \bC_{YX}\\
	&=\bC_X-\bC_X\bPi_\calA (\bPi_\calA(\bC_X+\bC_\calE)\bPi_\calA)^\dag \bPi_\calA \bC_X\\
	&:=\bC_X-\bC_0.
\end{align*}
Next, we compute $\tr(\bC_R)$, the main step of which is to deal with $\tr(\bC_0)$. By the operator trace definition in \cref{eq:operator-trace},
\begin{align}\label{eq:optr}
	\tr(\bC_0)&=\sum_{\tau=1}^{\infty}\sum_{k=1}^{n}\angles {\bC_0(\phi_k\otimes\psi_{\tau}),\phi_k\otimes\psi_{\tau}}\nn
	&=\sum_{\tau=1}^{\tau_0}\sum_{k=1}^{n} \langle(\bPi_\calA(\bC_X+\bC_\calE)\bPi_\calA)^\dag \bPi_\calA \bC_X(\phi_k\otimes\psi_{\tau}),\nn
	& \bPi_\calA \bC_X(\phi_k\otimes\psi_{\tau})\rangle.
\end{align}
To simplify this expression, we compute the elements in the inner products as follows:
\begin{align*}
	&\bPi_\calA \bC_X(\phi_k\otimes\psi_{\tau})=\\
	&p_X(k,\tau)\bPi_\calA(\phi_k\otimes\psi_{\tau})=p_X(k,\tau) \sum_{i\in\scU}\bPhi_{ik} v_i\otimes \psi_{\tau},\\
	&(\bPi_\calA(\bC_X+\bC_\calE)\bPi_\calA)^\dag \bPi_\calA \bC_X(\phi_k\otimes\psi_{\tau})=\\
	&p_X(k,\tau)\sum_{i\in\scU}\bPhi_{ik}(\bPi_\calA(\bC_X+\bC_\calE)\bPi_\calA)^\dag(v_i\otimes \psi_{\tau}).
\end{align*}
Note that $\bPi_\calA(\bC_X+\bC_\calE)\bPi_\calA=\bC_Y$, hence it is guaranteed to be compact and self-adjoint. To compute $\bC_Y^\dag(v_i\otimes \psi_{\tau})$, we utilize the result from \cref{lem:C_Ystructure}, which characterizes the restriction of $\bC_Y$ on $\widetilde{V_{\tau}}$ as an invertible matrix $\bO^{(\tau)}$. Therefore, $\sum_{j\in\scU}(\bO^{(\tau)-1})_{ij}v_j\otimes \psi_{\tau}$ is the preimage of $v_i\otimes \psi_{\tau}$. Besides, since it also belongs to $(\ker \bC_Y)^\perp$, we have $\bC_Y^\dag(v_i\otimes \psi_{\tau})=\sum_{j\in\scU}(\bO^{(\tau)-1})_{ij}v_j\otimes \psi_{\tau}$. Hence,
\begin{align*}
	&(\bPi_\calA(\bC_X+\bC_\calE)\bPi_\calA)^\dag \bPi_\calA \bC_X(\phi_k\otimes\psi_{\tau})\\
	&=p_X(k,\tau)\sum_{i\in\scU}\bPhi_{ik}\bC_Y^\dag(v_i\otimes \psi_{\tau})\\
	&=p_X(k,\tau)\sum_{i\in\scU}\sum_{j\in\scU}\bPhi_{ik}(\bO^{(\tau)-1})_{ij}v_j\otimes \psi_{\tau}.
\end{align*}
By substituting this result into each term in the double sum of \cref{eq:optr}, we obtain 
\begin{align*}
	&\angles{(\bPi_\calA(\bC_X+\bC_\calE)\bPi_\calA)^\dag \bPi_\calA \bC_X(\phi_k\otimes\psi_{\tau}),\bPi_\calA \bC_X(\phi_k\otimes\psi_{\tau})}\\
	&=p_X(k,\tau)^2\angles{\sum_{i\in\scU}\sum_{j\in\scU}\bPhi_{ik}(\bO^{(\tau)-1})_{ij}v_j\otimes \psi_{\tau},\sum_{l\in\scU}\bPhi_{lk} v_l\otimes \psi_{\tau}}\\
	&=p_X(k,\tau)^2\sum_{i\in\scU}\sum_{j\in\scU}\bPhi_{ik}(\bO^{(\tau)-1})_{ij}\bPhi_{jk}\\
	&=p_X(k,\tau)^2\bPhi_{\scU}\T(k,:)\bO^{(\tau)-1}\bPhi_{\scU}(:,k),
\end{align*}
where $\bPhi_{\scU}\T(k,:)$ denotes the $k$-th row of $\bPhi_{\scU}\T$.

Then, we have
\begin{align*}
	\tr(\bC_0)&=\sum_{\tau=1}^{\tau_0}\sum_{k=1}^{n} p_X(k,\tau)^2\bPhi_{\scU}\T(k,:)\bO^{(\tau)-1}\bPhi_{\scU}(:,k)\\
	&=\sum_{\tau=1}^{\tau_0}\sum_{k=1}^{n}p_X(k,\tau)^2\tr(\bO^{(\tau)-1}\bPhi_{\scU}(:,k)\bPhi_{\scU}\T(k,:))\\
	&=\sum_{\tau=1}^{\tau_0}\tr(\bO^{(\tau)-1}\sum_{k=1}^{n}p_X(k,\tau)^2\bPhi_{\scU}(:,k)\bPhi_{\scU}\T(k,:))\\
	&=\sum_{\tau=1}^{\tau_0}\tr((\bPhi_{\scU}\bLambda_\tau\bPhi_{\scU}\T)^{-1}\bPhi_{\scU}\bGamma_\tau\bPhi_{\scU}\T).
\end{align*}
Finally, we obtain the \gls{MSE} of the Wiener filter $\bG$ as
\begin{align*}
	\bbE\norm{R}^2 &= \tr(\bC_X)-\sum_{\tau=1}^{\tau_0}\tr((\bPhi_{\scU}\bLambda_\tau\bPhi_{\scU}\T)^{-1}\bPhi_{\scU}\bGamma_\tau\bPhi_{\scU}\T)\\
	&=\sum_{\tau=1}^{\tau_0}\sum_{k=1}^{n} p_X(k,\tau)-\sum_{\tau=1}^{\tau_0}\tr((\bPhi_{\scU}\bLambda_\tau\bPhi_{\scU}\T)^{-1}\bPhi_{\scU}\bGamma_\tau\bPhi_{\scU}\T).
\end{align*}
The proof is now complete.

\bibliographystyle{IEEEtran}
\bibliography{IEEEabrv,StringDefinitions,wholelib_v1}

\end{document}

%% file: preamble.tex
\usepackage[T1]{fontenc}
\usepackage[utf8]{inputenc}
\usepackage{amsmath,amssymb,amsfonts,mathrsfs,bm}
\usepackage{mathtools}
\usepackage{amsthm}
\usepackage{scalerel}
\usepackage{nicefrac}
\usepackage[shortlabels]{enumitem}
\usepackage{graphicx}
\usepackage{epstopdf}
\DeclareGraphicsExtensions{.eps,.png,.jpg,.pdf}

\usepackage{url}
\usepackage{colortbl}
\usepackage{booktabs}
\usepackage{multirow}
\usepackage[table,dvipsnames]{xcolor}
\usepackage[normalem]{ulem}
\usepackage{xparse}
\usepackage{calc}
\usepackage{etoolbox}

\makeatletter
\@ifpackageloaded{natbib}{
	\relax
}{
	\usepackage{cite}
}
\makeatother

\usepackage{array}
\newcolumntype{L}[1]{>{\raggedright\let\newline\\\arraybackslash\hspace{0pt}}m{#1}}
\newcolumntype{C}[1]{>{\centering\let\newline\\\arraybackslash\hspace{0pt}}m{#1}}
\newcolumntype{R}[1]{>{\raggedleft\let\newline\\\arraybackslash\hspace{0pt}}m{#1}}

\makeatletter
\let\MYcaption\@makecaption
\makeatother
\usepackage[font=footnotesize]{subcaption}
\makeatletter
\let\@makecaption\MYcaption
\makeatother

\usepackage{glossaries}
\makeatletter
\let\oldgls\gls
\let\oldglspl\glspl

\newcommand\fussy@ifnextchar[3]{%
	\let\reserved@d=#1%
	\def\reserved@a{#2}%
	\def\reserved@b{#3}%
	\futurelet\@let@token\fussy@ifnch}
\def\fussy@ifnch{%
	\ifx\@let@token\reserved@d
		\let\reserved@c\reserved@a
	\else
		\let\reserved@c\reserved@b
	\fi
	\reserved@c}

\renewcommand{\gls}[1]{%
\oldgls{#1}\fussy@ifnextchar.{\@checkperiod}{\@}}
\renewcommand{\glspl}[1]{%
\oldglspl{#1}\fussy@ifnextchar.{\@checkperiod}{\@}}

\newcommand{\@checkperiod}[1]{%
	\ifnum\sfcode`\.=\spacefactor\else#1\fi
}
\makeatother

\newacronym{wrt}{w.r.t.}{with respect to}
\newacronym{RHS}{R.H.S.}{right-hand side}
\newacronym{LHS}{L.H.S.}{left-hand side}
\newacronym{iid}{i.i.d.}{independent and identically distributed}

\usepackage{float}

\ifx\notloadhyperref\undefined
	\ifx\loadbibentry\undefined
		\usepackage[hidelinks,hypertexnames=false]{hyperref}
	\else
		\usepackage{bibentry}
		\makeatletter\let\saved@bibitem\@bibitem\makeatother
		\usepackage[hidelinks,hypertexnames=false]{hyperref}
		\makeatletter\let\@bibitem\saved@bibitem\makeatother
	\fi
\else
	\ifx\loadbibentry\undefined
		\relax
	\else
		\usepackage{bibentry}
	\fi
\fi

\usepackage[capitalize]{cleveref}
\crefname{equation}{}{}
\Crefname{equation}{}{}
\crefname{claim}{claim}{claims}
\crefname{step}{step}{steps}
\crefname{line}{line}{lines}
\crefname{condition}{condition}{conditions}
\crefname{dmath}{}{}
\crefname{dseries}{}{}
\crefname{dgroup}{}{}

\crefname{Problem}{Problem}{Problems}
\crefformat{Problem}{Problem~(#2#1#3)}
\crefrangeformat{Problem}{Problems~(#3#1#4) to~(#5#2#6)}

\crefname{Theorem}{Theorem}{Theorems}
\crefname{Corollary}{Corollary}{Corollaries}
\crefname{Proposition}{Proposition}{Propositions}
\crefname{Lemma}{Lemma}{Lemmas}
\crefname{Definition}{Definition}{Definitions}
\crefname{Example}{Example}{Examples}
\crefname{Assumption}{Assumption}{Assumptions}
\crefname{Remark}{Remark}{Remarks}
\crefname{Rem}{Remark}{Remarks}
\crefname{remarks}{Remarks}{Remarks}
\crefname{Appendix}{Appendix}{Appendices}
\crefname{Supplement}{Supplement}{Supplements}
\crefname{Exercise}{Exercise}{Exercises}
\crefname{Theorem_A}{Theorem}{Theorems}
\crefname{Corollary_A}{Corollary}{Corollaries}
\crefname{Proposition_A}{Proposition}{Propositions}
\crefname{Lemma_A}{Lemma}{Lemmas}
\crefname{Definition_A}{Definition}{Definitions}

\usepackage{crossreftools}
\ifx\notloadhyperref\undefined
\else
	\relax
\fi

\usepackage{algorithm,algorithmic}

\ifx\loadbreqn\undefined
	\relax
\else
	\usepackage{breqn}
\fi

\makeatletter
\def\cleartheorem#1{%
    \expandafter\let\csname#1\endcsname\relax
    \expandafter\let\csname c@#1\endcsname\relax
}
\def\clearthms#1{ \@for\tname:=#1\do{\cleartheorem\tname} }
\makeatother

\ifx\renewtheorem\undefined
	\ifx\useTheoremCounter\undefined
		\newtheorem{Theorem}{Theorem}
		\newtheorem{Corollary}{Corollary}
		\newtheorem{Proposition}{Proposition}
		\newtheorem{Lemma}{Lemma}
	\else
		\newtheorem{Theorem}{Theorem}
		
		\newtheorem{Proposition}[Theorem]{Proposition}
	\fi

	\newtheorem{Definition}{Definition}
	\newtheorem{Example}{Example}
	
	\newtheorem{Assumption}{Assumption}

\fi

\theoremstyle{remark}

\theoremstyle{plain}

\newcommand{\qednew}{\nobreak \ifvmode \relax \else
		\ifdim\lastskip<1.5em \hskip-\lastskip
			\hskip1.5em plus0em minus0.5em \fi \nobreak
		\vrule height0.75em width0.5em depth0.25em\fi}



\newcommand{\nn}{\nonumber\\ }

\NewDocumentCommand{\movedownsub}{e{^_}}{%
	\IfNoValueTF{#1}{%
		\IfNoValueF{#2}{^{}}
	}{%
		^{#1}
	}%
	\IfNoValueF{#2}{_{#2}}
}

\let\latexchi\chi
\RenewDocumentCommand{\chi}{}{\latexchi\movedownsub}

\newcommand{\Real}{\mathbb{R}}

\newcommand{\Complex}{\mathbb{C}}

\newcommand{\iu}{\mathfrak{i}\mkern1mu}

\newcommand{\calA}{\mathcal{A}}
\newcommand{\calB}{\mathcal{B}}
\newcommand{\calC}{\mathcal{C}}

\newcommand{\calE}{\mathcal{E}}
\newcommand{\calF}{\mathcal{F}}
\newcommand{\calG}{\mathcal{G}}
\newcommand{\calH}{\mathcal{H}}

\newcommand{\calN}{\mathcal{N}}

\newcommand{\bA}{\mathbf{A}}

\newcommand{\bB}{\mathbf{B}}

\newcommand{\bC}{\mathbf{C}}

\newcommand{\bD}{\mathbf{D}}

\newcommand{\bG}{\mathbf{G}}

\newcommand{\bI}{\mathbf{I}}

\newcommand{\bK}{\mathbf{K}}

\newcommand{\bL}{\mathbf{L}}

\newcommand{\bO}{\mathbf{O}}

\newcommand{\bT}{\mathbf{T}}

\newcommand{\bW}{\mathbf{W}}

\newcommand{\bX}{\mathbf{X}}

\newcommand{\bbC}{\mathbb{C}}

\newcommand{\bbE}{\mathbb{E}}

\newcommand{\bbN}{\mathbb{N}}

\newcommand{\bbP}{\mathbb{P}}

\newcommand{\scE}{\mathscr{E}}
\newcommand{\scF}{\mathscr{F}}
\newcommand{\scG}{\mathscr{G}}
\newcommand{\scH}{\mathscr{H}}

\newcommand{\scU}{\mathscr{U}}
\newcommand{\scV}{\mathscr{V}}

\DeclareSymbolFont{bsfletters}{OT1}{cmss}{bx}{n}
\DeclareSymbolFont{ssfletters}{OT1}{cmss}{m}{n}
\DeclareMathSymbol{\bsfGamma}{0}{bsfletters}{'000}
\DeclareMathSymbol{\ssfGamma}{0}{ssfletters}{'000}
\DeclareMathSymbol{\bsfDelta}{0}{bsfletters}{'001}
\DeclareMathSymbol{\ssfDelta}{0}{ssfletters}{'001}
\DeclareMathSymbol{\bsfTheta}{0}{bsfletters}{'002}
\DeclareMathSymbol{\ssfTheta}{0}{ssfletters}{'002}
\DeclareMathSymbol{\bsfLambda}{0}{bsfletters}{'003}
\DeclareMathSymbol{\ssfLambda}{0}{ssfletters}{'003}
\DeclareMathSymbol{\bsfXi}{0}{bsfletters}{'004}
\DeclareMathSymbol{\ssfXi}{0}{ssfletters}{'004}
\DeclareMathSymbol{\bsfPi}{0}{bsfletters}{'005}
\DeclareMathSymbol{\ssfPi}{0}{ssfletters}{'005}
\DeclareMathSymbol{\bsfSigma}{0}{bsfletters}{'006}
\DeclareMathSymbol{\ssfSigma}{0}{ssfletters}{'006}
\DeclareMathSymbol{\bsfUpsilon}{0}{bsfletters}{'007}
\DeclareMathSymbol{\ssfUpsilon}{0}{ssfletters}{'007}
\DeclareMathSymbol{\bsfPhi}{0}{bsfletters}{'010}
\DeclareMathSymbol{\ssfPhi}{0}{ssfletters}{'010}
\DeclareMathSymbol{\bsfPsi}{0}{bsfletters}{'011}
\DeclareMathSymbol{\ssfPsi}{0}{ssfletters}{'011}
\DeclareMathSymbol{\bsfOmega}{0}{bsfletters}{'012}
\DeclareMathSymbol{\ssfOmega}{0}{ssfletters}{'012}

\newcommand{\bdelta}{\bm{\delta}}

\newcommand{\bGamma}{\bm{\Gamma}}
\newcommand{\bLambda}{\bm{\Lambda}}

\newcommand{\bPsi}{\bm{\Psi}}

\newcommand{\bPhi}{\bm{\Phi}}
\newcommand{\bPi}{\bm{\Pi}}

\newcommand{\hbPsi}{\widehat{\bPsi}}

\makeatletter
\newcommand*\rel@kern[1]{\kern#1\dimexpr\macc@kerna}
\newcommand*\widebar[1]{%
  \begingroup
  \def\mathaccent##1##2{%
    \rel@kern{0.8}%
    \overline{\rel@kern{-0.8}\macc@nucleus\rel@kern{0.2}}%
    \rel@kern{-0.2}%
  }%
  \macc@depth\@ne
  \let\math@bgroup\@empty \let\math@egroup\macc@set@skewchar
  \mathsurround\z@ \frozen@everymath{\mathgroup\macc@group\relax}%
  \macc@set@skewchar\relax
  \let\mathaccentV\macc@nested@a
  \macc@nested@a\relax111{#1}%
  \endgroup
}
\makeatother

\DeclareMathOperator*{\argmin}{arg\,min}

\DeclareMathOperator{\diag}{diag}

\DeclareMathOperator{\tr}{tr}

\DeclareMathOperator{\spn}{span}

\DeclareMathOperator{\var}{var}

\DeclareMathOperator{\cov}{cov}

\DeclareMathOperator{\vect}{vec}

\DeclareMathOperator{\ima}{im}

\newcommand{\ifbcdot}[1]{\ifblank{#1}{\cdot}{#1}}

\DeclarePairedDelimiterX\abs[1]{\lvert}{\rvert}{\ifbcdot{#1}}
\DeclarePairedDelimiterX\parens[1]{(}{)}{\ifbcdot{#1}}
\DeclarePairedDelimiterX\brk[1]{[}{]}{\ifbcdot{#1}}
\DeclarePairedDelimiterX\braces[1]{\{}{\}}{\ifbcdot{#1}}
\DeclarePairedDelimiterX\angles[1]{\langle}{\rangle}{\ifblank{#1}{\cdot,\cdot}{#1}}
\DeclarePairedDelimiterX\ip[2]{\langle}{\rangle}{\ifbcdot{#1},\ifbcdot{#2}}
\DeclarePairedDelimiterX\norm[1]{\lVert}{\rVert}{\ifbcdot{#1}}
\DeclarePairedDelimiterX\ceil[1]{\lceil}{\rceil}{\ifbcdot{#1}}
\DeclarePairedDelimiterX\floor[1]{\lfloor}{\rfloor}{\ifbcdot{#1}}

\DeclarePairedDelimiterXPP\trace[1]{\operatorname{Tr}}{(}{)}{}{\ifbcdot{#1}} 
\DeclarePairedDelimiterXPP\col[1]{\operatorname{col}}{\{}{\}}{}{\ifbcdot{#1}} 
\DeclarePairedDelimiterXPP\row[1]{\operatorname{row}}{\{}{\}}{}{\ifbcdot{#1}} 
\DeclarePairedDelimiterXPP\erf[1]{\operatorname{erf}}{(}{)}{}{\ifbcdot{#1}}
\DeclarePairedDelimiterXPP\erfc[1]{\operatorname{erfc}}{(}{)}{}{\ifbcdot{#1}}
\DeclarePairedDelimiterXPP\KLD[2]{D}{(}{)}{}{\ifbcdot{#1}\, \delimsize\|\, \ifbcdot{#2}} 
\DeclarePairedDelimiterXPP\op[2]{\operatorname{#1}}{(}{)}{}{#2} 

\newcommand{\convas}{\stackrel{\mathrm{a.s.}}{\longrightarrow}}

\newcommand{\T}{^{\intercal}}
\newcommand{\setcomp}{^{\mathsf{c}}} 
\newcommand{\ud}{\,\mathrm{d}} 

\newcommand{\bzero}{\bm{0}}

\DeclarePairedDelimiterXPP\indicate[1]{{\bf 1}}{\{}{\}}{}{\ifbcdot{#1}}
\newcommand{\indicator}[1]{{\bf 1}_{\braces*{\ifbcdot{#1}}}}

\newcommand{\ofrac}[1]{{\frac{1}{#1}}}

\providecommand\given{}

\DeclarePairedDelimiterX\Set[2]\{\}{%
\renewcommand\given{\SetSymbol[\delimsize]{#1}}
#2
}
\DeclarePairedDelimiterX\Setc[1]\{\}{%
\renewcommand\given{\SetSymbol{:}}
#1
}

\NewDocumentCommand\set{s o m}{%
	\IfBooleanTF#1%
	{\IfValueTF{#2}{\Set*{#2}{#3}}{\Setc*{#3}}}%
	{\IfValueTF{#2}{\Set{#2}{#3}}{\Setc{#3}}}%
}

\NewDocumentCommand{\evalat}{ s O{\big} m e{_^} }{%
\IfBooleanTF{#1}%
{\left. #3 \right|}{#3#2|}%
\IfValueT{#4}{_{#4}}%
\IfValueT{#5}{^{#5}}%
}

\NewDocumentCommand \ifcondp {m m} {%
	#1%
	\IfValueT{#2}{\given #2}%
}

\providecommand\given{}
\DeclarePairedDelimiterXPP\cprob[1]{}(){}{
\renewcommand\given{\nonscript\,\delimsize\vert\allowbreak\nonscript\,\mathopen{}}
\ifcondp#1
}
\DeclarePairedDelimiterXPP\cexp[1]{}[]{}{
\renewcommand\given{\nonscript\,\delimsize\vert\allowbreak\nonscript\,\mathopen{}}
\ifcondp#1
}

\DeclareDocumentCommand \P { s e{_^} >{\SplitArgument{ 1 }{ @| }}d() g } {%
	\mathbb{P}%
	\IfBooleanTF{#1}%
		{
			\IfValueT{#2}{_{#2}}%
			\IfValueT{#3}{^{#3}}%
			\IfValueTF{#5}%
				{\cprob{#4 \given #5}}%
				{\IfValueT{#4}{\cprob{#4}}}%
		}%
		{
			\IfValueT{#2}{_{#2}}%
			\IfValueT{#3}{^{#3}}%
			\IfValueTF{#5}%
				{\cprob*{#4 \given #5}}%
				{\IfValueT{#4}{\cprob*{#4}}}%
		}%
}

\DeclareDocumentCommand \E { s e{_^} >{\SplitArgument{ 1 }{ @| }}d[] g } {%
	\mathbb{E}%
	\IfBooleanTF{#1}%
		{
			\IfValueT{#2}{_{#2}}%
			\IfValueT{#3}{^{#3}}%
			\IfValueTF{#5}%
				{\cexp{#4 \given #5}}%
				{\IfValueT{#4}{\cexp{#4}}}%
		}%
		{
			\IfValueT{#2}{_{#2}}%
			\IfValueT{#3}{^{#3}}%
			\IfValueTF{#5}%
				{\cexp*{#4 \given #5}}%
				{\IfValueT{#4}{\cexp*{#4}}}%
		}%
}

\ExplSyntaxOn
\NewDocumentCommand \dist {m o o} {%
\mathrm{#1}\left(%
	\IfValueT{#3}{%
		\tl_if_blank:nTF{ #3 }{\cdot\, \middle|\, }{#3\, \middle|\, }%
	}
	\IfValueT{#2}{#2}%
\right)%
}
\ExplSyntaxOff

\NewDocumentCommand {\cbrace} {t+ D[]{black} D(){\widthof{#5}} m m } {%
	\begingroup%
		\color{#2}
		\IfBooleanTF{#1}{%
			\overbrace{#4}^%
		}{
			\underbrace{#4}_%
		}%
		{\parbox[c]{#3}{\centering\footnotesize{#5}}}%
	\endgroup%
}

\let\oldforall\forall
\renewcommand{\forall}{\oldforall \, }

\let\oldexist\exists
\renewcommand{\exists}{\oldexist \, }

\graphicspath{{./Figures/}{./figures/}}
\DeclareDocumentCommand{\includeCroppedPdf}{ o O{./Figures/} m }{
	\IfFileExists{#2#3-crop.pdf}{}{%
		\immediate\write18{pdfcrop #2#3.pdf #2#3-crop.pdf}}%
	\includegraphics[#1]{#2#3-crop.pdf}
}

\makeatletter
\newcommand*{\addFileDependency}[1]{
  \typeout{(#1)}
  \@addtofilelist{#1}
  \IfFileExists{#1}{}{\typeout{No file #1.}}
}
\makeatother

\definecolor{gray90}{gray}{0.9}

\ifx\nohighlights\undefined

	\newcommand{\msout}[1]{\text{\color{green} \sout{\ensuremath{#1}}}}
	\newcommand{\del}[1]{{\color{green}\ifmmode \msout{#1}\else\sout{#1}\fi}}
\else

	\newcommand{\msout}[1]{#1}
	\newcommand{\del}[1]{#1}
\fi

\newcommand{\hhide}[1]{}

\ifx\diagnoselabel\undefined
	\relax
\else
	\makeatletter
	\def\@testdef #1#2#3{%
		\def\reserved@a{#3}\expandafter \ifx \csname #1@#2\endcsname
			\reserved@a  \else
			\typeout{^^Jlabel #2 changed:^^J%
				\meaning\reserved@a^^J%
				\expandafter\meaning\csname #1@#2\endcsname^^J}%
			\@tempswatrue \fi}
	\makeatother
\fi
